%% file: ArxivVer2.tex
\documentclass[11pt]{article}
\usepackage{jheppub} 

\usepackage[utf8]{inputenc}
\usepackage{amsmath, amsthm, amsfonts, mathtools}
\usepackage{hyperref}
\usepackage{tabularray}

\usepackage{algorithmic}
\usepackage{textcomp}
\usepackage{comment}
\usepackage{xcolor}
\usepackage{url}
\usepackage{csquotes}
\usepackage{float}
\usepackage{natbib}
\def\BibTeX{{\rm B\kern-.05em{\sc i\kern-.025em b}\kern-.08em
    T\kern-.1667em\lower.7ex\hbox{E}\kern-.125emX}}
    \title{Covariant Carrollian Electric and Magnetic Limits of General Relativity}

\author[a]{Tanmay Patil,}
\author[b]{S. Shankaranarayanan}

\affiliation[a]{Department of Physics, Indian Institute of Technology Bombay, Mumbai}
\affiliation[b]{Department of Physics, Indian Institute of Technology Bombay, Mumbai}

\emailAdd{210260058@iitb.ac.in}
\emailAdd{shanki@iitb.ac.in}

\abstract{
The Carrollian limit ($c \to 0$) of General Relativity provides the geometric language for describing null hypersurfaces, such as black hole event horizons and null infinity. Motivated by the well-established electric and magnetic limits of Galilean electromagnetism, we perform a systematic analysis of the low-velocity limit of linearized gravity to derive its Carrollian counterparts. Using a 1+3 covariant decomposition, we study the transformation properties of linear tensor perturbations (gravitational waves) on a Friedmann-Lemaître-Robertson-Walker background under Carrollian boosts. We demonstrate that, analogous to the electromagnetic case, the full set of linearized Einstein's equations is not Carrollian-invariant. Instead, the theory bifurcates into two distinct and consistent frameworks: a \emph{Carrollian Electric Limit} and a \emph{Carrollian Magnetic Limit}. In the electric limit, dynamics are frozen, leaving a static theory of tidal forces ($E_{ab}$) constrained by the matter distribution. In contrast, the \emph{Magnetic Limit} yields a consistent dynamical theory where the magnetic part of the Weyl tensor ($H_{ab}$), which governs gravito-magnetic and radiative effects, remains well-defined and is sourced by the spacetime shear. This framework resolves ambiguities in defining Carrollian gravity and provides a robust theory for gravito-magnetic dynamics in ultra-relativistic regimes. Our results have direct implications for the study of black hole horizons, gravitational memory, and the holographic principle.
}

\begin{document}

\maketitle

\begin{abstract}
~~~~
\end{abstract}

\section{Introduction}

The principle of Lorentz invariance lies at the heart of modern physics, uniting space and time into a single entity --- spacetime. Yet, some of the most pressing questions in physics emerge in regimes where this very structure is pushed to its limits. Understanding physics in extreme environments, such as near black hole horizons or in the very early universe, requires us to explore novel theoretical frameworks. A powerful strategy is to study singular limits of General Relativity (GR), where the spacetime symmetries are deformed. The two most prominent examples are the \emph{Galilean limit} ($c \to \infty$), which recovers non-relativistic physics~\cite{DauntCourt:1990abc,Dautcourt:1997hb}, and the \emph{Carrollian limit} ($c \to 0$), which describes an ``ultra-local" world where light cones collapse into lines, and points become causally disconnected~\cite{Duval:1984cj,Duval:1990hj}.

The connection between the Carrollian limit and null surfaces is fundamental, not merely an analogy. On any null hypersurface, the metric becomes degenerate along the direction of propagation. This degeneracy forces causal signals to be confined to the null generators, completely forbidding communication across them in the transverse directions. This physical situation—where causality is restricted and propagation is channeled along one dimension—is a direct manifestation of the ``ultra-local" causal structure inherent to Carrollian geometry. Therefore, the Carrollian framework is not just a useful approximation but the natural and intrinsic language to describe dynamics on \emph{null hypersurfaces} --- surfaces like black hole event horizons~\cite{Donnay:2015abr,Donnay:2019jiz}, cosmological horizons~\cite{Campoleoni:2022ebj}, and the celestial sphere at null infinity~\cite{Hansen:2021fxi}. This realization has ignited a surge of interest, connecting Carrollian geometry to foundational topics like the black hole information paradox, the nature of gravitational entropy, and the holographic principle (BMS symmetries and celestial holography)~\cite{Duval:2014uoa,Hartong:2015xda}. While the dynamics of Carrollian fluids and electromagnetic fields are now relatively well-understood~\cite{Ciambelli:2018xat}, a complete and consistent description of Carrollian gravity remains an open frontier. This work addresses a key piece of that puzzle.

A direct translation of limiting procedures from electromagnetism to gravity is fraught with challenges~\cite{Hansen:2021fxi}. In electromagnetism, the electric ($c \to 0$) and magnetic ($c \to 0$) limits are distinct, yielding different physical theories that do not transform covariantly under a single set of rules~\cite{Levy-Leblond:1965dsc,LeBellac:1973koc,Heras_2010}. A similar discord appears in gravity: the Weyl conformal curvature tensor, $C_{abcd}$, can be decomposed into its electric and magnetic parts. These two components represent the independent degrees of freedom of the free gravitational field~\cite{1996-vanElst.Ellis-CQG,2012-ellis_maartens_maccallum-Book}: Specifically, the Electric Weyl Tensor ($E_{ab}$) represents the tidal forces of the gravitational field. This includes both static tidal distortions, such as those around a star or black hole, and the time-varying tidal strains that are the physical manifestation of a passing gravitational wave detected by instruments like LIGO. The Magnetic Weyl Tensor ($H_{ab}$) represents the gravito-magnetic aspects of the field. This includes effects like frame-dragging caused by mass currents and rotation. In the context of gravitational waves, it describes the complementary radiative degrees of freedom that, together with $E_{ab}$, fully characterize the propagating wave.

The standard tools for studying gravitational dynamics, such as the \emph{3+1 (ADM) Cauchy formulation}, are inherently designed for predictable evolution from an initial \emph{spacelike} hypersurface~\cite{Frittelli:1994stb}.  This framework fundamentally breaks down when one attempts to formulate it on a null surface like a black hole horizon, where the normal vector is lightlike and a well-posed initial value problem cannot be defined. Since, Carrollian geometry provides the correct intrinsic language for these null boundaries, it is crucial that the ``3+1" structure of a Carrollian theory is physically distinct from a Lorentzian Cauchy evolution.  
This structure circumvents the issues that plague the standard initial value formulation on a null surface, allowing for a consistent description of the dynamics \emph{on} the boundary itself. Our approach is built on this insight, using the 1+3 and 1+2+1 covariant formalisms to capture these Carrollian features without ambiguity.
While adaptations like the \emph{1+2+1 null decomposition} exist \cite{Mars:1993mj}, they introduce their own complexities, especially when defining conserved quantities and propagation equations for gravitational radiation in these degenerate backgrounds~\cite{Petkou:2022bmz}.

While the non-relativistic limit of General Relativity is often equated with Newtonian gravity, the underlying structure of spacetime admits more exotic possibilities. 
The Galilean limit ($c \to \infty$) establishes an absolute sense of time and instantaneous propagation, forming the bedrock of Newtonian mechanics.
The Carrollian limit ($c \to 0$), in contrast, leads to a world where light cones collapse into lines. This results in an absolute sense of space where causal contact is lost. This latter limit, named for Lewis Carroll's  ``Through the Looking-Glass" seemingly paradoxical world, was long considered a curiosity but is now understood to be the essential framework for describing physics on null hypersurfaces like black hole horizons. 
This subtlety is well-established in electromagnetism. As first shown by Lévy-Leblond, Sengupta and Le Bellac, applying a low-velocity limit under different assumptions about the dominant fields splits Maxwell's theory into two distinct non-relativistic frameworks: the well-known electric and magnetic limits~\cite{Levy-Leblond:1965dsc,SenGupta:1966abc,LeBellac:1973koc}. This raises a crucial question that is the central motivation for our work: Can a similar, physically consistent structure be found for both electromagnetism and gravity under Carrollian transformations? or What is the analogous structure of General Relativity in the Carrollian limit, and can we define consistent dynamics for its gravitational degrees of freedom?

In this work, we systematically investigate the \emph{Magnetic Carrollian Limit of GR} to establish a consistent dynamical framework for gravitational degrees of freedom in the $c \to 0$ regime. Our focus is placed specifically on the magnetic limit because, as our analysis will demonstrate, this is the regime that preserves non-trivial, dynamical degrees of freedom on the null surface. In contrast, the Electric Carrollian Limit is found to be far more restrictive, collapsing to a static theory with no propagating modes, making it less suitable for studying dynamic phenomena like horizon evolution or gravitational memory effects. Our central goal is to resolve the inconsistencies that arise from the non-uniform scaling of the gravitational field and build a unified, covariant framework.
To do this, we employ a \emph{1+3 covariant decomposition} formalism, pioneered by Ellis and others, which is manifestly independent of coordinate choices and adaptable to different observer congruences~\cite{1998-Ellis-NATOSci,1996-vanElst.Ellis-CQG,2012-ellis_maartens_maccallum-Book}. We perform our analysis on a Friedmann-Lema\^itre-Robertson-Walker (FLRW) cosmological background, allowing us to study linear tensor perturbations in a controlled setting~\cite{1989-Ellis.Bruni-PRD,1989-Ellis.Hwang.Bruni-PRD}.

To clarify the novel contributions of our approach, we first distinguish it from foundational studies of Carrollian gravity, notably by Henneaux~\cite{Henneaux:1979vn}, Salgado-Rebolledo \cite{Henneaux:2021gwp} and Campoleoni et al. \cite{Campoleoni:2022ebj}. These authors have established the existence of two inequivalent electric and magnetic limits. Their approaches, rooted in the Hamiltonian formalism and the gauging of the Carroll algebra, derive these limits by analyzing the structure of the action principle. Our work provides a complementary perspective by investigating these limits from a manifestly covariant and geometric viewpoint. Using the 1+3 decomposition of spacetime, we derive the limiting theories directly at the level of the gravitational field equations. This method allows us to explicitly track the surviving dynamical content, revealing that while the electric limit collapses to a static, constraint-dominated system, the magnetic limit preserves non-trivial dynamics for the magnetic part of the Weyl tensor. Our analysis therefore complements the Hamiltonian and algebraic treatments by offering a covariant formulation that illuminates the geometric and dynamical nature of these Carrollian theories.

{Finally, we clarify the methodology of our approach. Our analysis in this work is performed at the level of the \emph{covariant field equations}, not the action principle. As is well-known, these two procedures are not necessarily equivalent in singular limits. For instance, in the Galilean limit, the correct Newton-Cartan equations of motion do not follow from a simple limit of the Einstein-Hilbert action. Deriving a consistent action for Carrollian gravity is a highly non-trivial task, often requiring a gauging of the Carroll algebra. Our work is complementary to those efforts. By focusing on the invariance of the equations of motion, we provide a direct, physical, and geometric picture of the surviving dynamical degrees of freedom, which can serve as a guide for what any complete action-based formulation must reproduce.}

Specifically, our key result is the demonstration that a consistent, dynamical theory for the \emph{magnetic part of the Weyl tensor} survives the $c \to 0$ limit. Remarkably, the resulting field equations are structurally analogous to their counterparts in the well-defined magnetic limit of electromagnetism. By providing a unified treatment for both timelike and null congruences, we show that gravitational radiation remains a well-defined and propagating phenomenon within this framework. This work lays the foundation for resolving key issues in Carrollian holography, gravitational memory effects, and the fundamental physics governing black hole horizons.

The work is organized as follows: In Sec.~\eqref{1+3}, we discuss the $1+3$ covariant formalism. Since this covariant approach is less common in some literature, we provide the essential details of the formalism here. Readers already familiar with the 1+3 covariant framework can proceed to Sec.\eqref{sec:NullFormalism} without loss of continuity. Sec.~\eqref{sec:NullFormalism} extends the $1 + 3$ covariant formalism for light-like particles and obtains the Maxwell and Gravity equations for arbitrary space-time. In Sec.~\eqref{sec:GalileanCarrollian}, we discuss different Galilean and Carrollian limits of space-time. Secs.~\ref{sec:Galilean-EM},\ref{sec:Instat-EM}, \ref{sec:Carroll-EM} discuss the different Galilean, Instantaneous and Carrollian limits of Maxwell's electrodynamics.  Secs.~\ref{sec:Galiliean-Grav}, \ref{sec:Carroll-Grav} discuss the different Galilean, and Carrollian limits of gravity. We discuss implications, key results and discuss
future directions in Sec.~\eqref{sec:Conc}. The  two appendices contain details of the calculations presented in the main text. 

We use the metric signature $(-,+,+,+)$. Latin alphabets denote the 4-dimensional space-time coordinates. An overdot represents the 
covariant derivative along the observer worldline. We set $G =1$. 
 
\section{1+3 covariant formalism} \label{1+3}

While General Relativity is fundamentally a 4-dimensional covariant theory, a clear physical interpretation often requires separating spacetime into ``space" and ``time" from a particular observer's perspective. The \emph{1+3 covariant formalism} is the ideal tool for this, as it allows for such a split without sacrificing coordinate independence~\cite{1996-vanElst.Ellis-CQG,1998-Ellis-NATOSci}. This is achieved by selecting a preferred family of observer, followed by spliting all physical quatity into temporal and spatial componets. The temporal componet is obtained by projecting the quatity along the observer and the spatial components lie on the surface orthogonal to the tangent to the observer.
This procedure systematically recasts the 4D Einstein Field Equations into a more intuitive set of 3D \emph{evolution equations}, which dictate the system's dynamics, and \emph{constraint equations}, which must be satisfied at any given moment. This framework thus provides a powerful and physically transparent way to analyze gravitational systems.

\subsection{Projectors and Tensor Decomposition}

In the cosmological context, we can always define a preferred timelike congruence representing the average flow of cosmic matter. These worldlines are followed by \emph{fundamental observers} with a 4-velocity vector field $u^a$, normalized such that:
\begin{equation}
    u^a = \frac{dx^a}{d\tau}, \qquad u_a u^a = -1,
    \label{eq:4velocity}
\end{equation}
where $\tau$ is the proper time measured along the observer's worldline.
Given the observer field $u^a$, we can uniquely decompose any tensor into parts parallel and orthogonal to the direction of time. This is accomplished using two fundamental projection operators:
\begin{enumerate}
    \item The tensor $U^a{}_b = -u^a u_b$ projects vectors \emph{parallel} to the observer's 4-velocity.
\item The tensor $h_{ab} = g_{ab} + u_a u_b$ is the projection tensor that projects quantities onto the 3D surface. As we show later, spacetimes with no vorticity (like, FLRW) $h_{ab}$ becomes metric in the 3D surface.
\end{enumerate}
These operators satisfy the expected properties of projectors:
\begin{equation}
h^a{}_c h^c{}_b = h^a{}_b, \qquad h_{ab} u^b = 0, \qquad h^a{}_a = 3
\end{equation}
Using them, any vector $V^a$ can be split into its temporal and spatial parts: 
\begin{equation}
V^a = (-V_b u^b)u^a + h^a{}_c V^c \, . 
\end{equation}
The geometry of the 3-D rest-spaces is also equipped with a spatial volume element, defined by contracting the 4D volume element $\eta_{abcd}$ with the observer's 4-velocity:
$$\eta_{abc} = u^d \eta_{dabc} \quad \Rightarrow \quad \eta_{abc} = \eta_{[abc]}, \quad \eta_{abc} u^c = 0$$
where $\eta_{0123} = \sqrt{-g}$.

\subsubsection{Covariant Derivatives}
To study dynamics, we need derivative operators that respect this 1+3 split~\cite{1998-Ellis-NATOSci}.
\begin{enumerate}
\item \underline{\it  Temporal Derivative:} The rate of change of any tensor $T$ as measured by a fundamental observer is the covariant derivative along their worldline. We denote this with an overdot:
\begin{equation}
    \dot{T}^{ab\dots}{}_{cd\dots} \equiv u^e \nabla_e T^{ab\dots}{}_{cd\dots}
\end{equation}
\item \underline{\it  Spatial Derivative:} The spatial gradient is defined by projecting the 4-D covariant derivative onto the observer's rest space. For a tensor $T$, this is:
\begin{equation}
\widetilde{\nabla}_e T^{ab\dots}{}_{cd\dots} \equiv h^a{}_f h^b{}_g \dots h^p{}_c h^q{}_d \dots h^r{}_e \nabla_r T^{fg\dots}{}_{pq\dots}
\end{equation}
    Here, the projection $h$ is applied to the derivative index and all free indices of the tensor $T$, ensuring the resulting object is purely spatial. It's important to note that if the observer congruence has non-zero vorticity ($\omega_{ab} \neq 0$), the commutator of these spatial derivatives does not vanish, meaning $\widetilde{\nabla}_a$ is not a true 3-dimensional Christoffel connection.
\end{enumerate}

\subsubsection{Irreducible Spatial Decompositions}

Finally, we introduce a notation for the irreducible parts of spatial tensors. The angled brackets $\langle \dots \rangle$ denote the \emph{Projected Symmetric Trace-Free (PSTF)} part of a tensor. This operation isolates the shear or anisotropic part of a physical quantity.

For any spatial vector $v^a$ (where $v^a u_a = 0$), the projection is trivial:
$$v^{\langle a \rangle} \equiv h^a{}_b v^b = v^a$$
For a second-rank spatial tensor $T^{ab}$, the PSTF part is constructed by making it symmetric and removing its trace:
\begin{equation}
T^{\langle ab \rangle} \equiv \left[ h^{(a}{}_c h^{b)}{}_d - \frac{1}{3} h^{ab} h_{cd} \right] T^{cd}
\end{equation}
This notation also applies to time derivatives of spatial tensors. For example, the \emph{Fermi derivative} of a spatial vector $v^a$ is its projected time derivative, $\dot{v}^{\langle a \rangle} = h^a{}_b \dot{v}^b$, which represents the rate of change of the vector as seen by the observer, with any component generated along $u^a$ projected out.

\subsection{Kinematical Decomposition of the Observer Congruence}

The 1+3 formalism allows us to describe the geometry of a fluid flow by examining the relative motion of nearby fundamental observers. This is encoded in the covariant derivative of the 4-velocity field, $\nabla_b u_a$, which can be irreducibly decomposed as:
\begin{equation}
\nabla_b u_a = -u_b \dot{u}_a + \frac{1}{3}\Theta h_{ab} + \sigma_{ab} + \omega_{ab} \, .
\end{equation}
Each term has a distinct physical interpretation:
\begin{enumerate}
\item \underline{\it  Acceleration ($\dot{u}_a = u^b \nabla_b u_a$):} This vector describes the acceleration of the observers. It's non-zero if the worldlines are not geodesics, implying the presence of non-gravitational forces.

\item \underline{\it  Expansion Scalar ($\Theta = \widetilde{\nabla}_a u^a$):} This scalar measures the isotropic rate of change in the volume of the fluid element. A positive $\Theta$ signifies expansion, while a negative $\Theta$ signifies contraction. In cosmology, the \emph{Hubble scalar} is defined as $H = \Theta/3$.

\item \underline{\it  Shear Tensor ($\sigma_{ab} = \widetilde{\nabla}_{\langle b} u_{a \rangle}$):} This symmetric, trace-free tensor describes the rate of distortion of the fluid element at constant volume. For instance, it quantifies how an initially spherical volume of observers is distorted into an ellipsoid. It represents anisotropic expansion. It satisfies:
    $$\sigma_{ab} = \sigma_{(ab)}, \qquad \sigma^a{}_a = 0, \qquad \sigma_{ab} u^b = 0$$

\item \underline{\it  Vorticity Tensor ($\omega_{ab} = \widetilde{\nabla}_{[b} u_{a]}$):} This anti-symmetric tensor measures the average angular velocity of the fluid element—its rotation relative to a non-rotating (Fermi-transported) frame. It satisfies:
    $$\omega_{ab} = \omega_{[ab]}, \qquad \omega_{ab} u^b = 0$$
\end{enumerate}

\subsubsection{1+3 Decomposition of the Energy-Momentum Tensor}

Similarly, the matter content of the universe, described by the energy-momentum tensor $T_{ab}$, can be decomposed from the perspective of the observers $u^a$. This reveals the physical properties of the cosmic fluid as measured in its own rest frame. The general decomposition is:
\begin{equation}
T_{ab} = \mu u_a u_b + q_a u_b + u_a q_b + p h_{ab} + \pi_{ab}
\end{equation}
The components are defined by projecting $T_{ab}$ parallel and orthogonal to $u^a$:
\begin{enumerate}
    \item \underline{\it  Energy Density ($\mu = T_{ab} u^a u^b$):} The energy density as measured by a fundamental observer.

\item \underline{\it  Energy Flux ($q_a = -T_{bc} u^b h^c{}_a$):} The flow of energy (heat) within the observer's 3D rest space. It represents energy transport relative to the bulk motion of the fluid. It is purely spatial, so $q_a u^a = 0$.

\item \underline{\it  Isotropic Pressure ($p = \frac{1}{3} T_{ab} h^{ab}$)}: The standard pressure that acts equally in all directions within the fluid's rest frame.

\item \underline{\it  Anisotropic Stress ($\pi_{ab} = T_{cd} h^c{}_{\langle a} h^d{}_{b \rangle}$)}: The trace-free part of the spatial stress tensor. It represents internal friction or viscosity—stresses that are not uniform in all directions. It satisfies:
    $$\pi_{ab} = \pi_{(ab)}, \qquad \pi^a{}_a = 0, \qquad \pi_{ab} u^b = 0$$
\end{enumerate}

\subsubsection{Physical Constraints and Equations of State}

The physics of the cosmic medium is determined by the relationships between these quantities. The simplest and most common model is that of a \emph{perfect fluid}, which is defined by the absence of heat flow and anisotropic stresses:
$$q_a = 0, \qquad \pi_{ab} = 0 \quad \implies \quad T_{ab} = (\mu + p) u_a u_b + p g_{ab}$$
For a perfect fluid, the system is closed by an \emph{equation of state}, typically of the form $p=p(\mu)$. A pressure-free perfect fluid ($p=0$) is often used to model dust or \emph{Cold Dark Matter (CDM)}.

For any realistic matter model, certain physical constraints must be imposed. \emph{Energy conditions}, such as the Weak Energy Condition ($\mu \geq 0$, $\mu+p \geq 0$) and the Strong Energy Condition ($\mu+3p \geq 0$), ensure physically reasonable behavior. Furthermore, to ensure causality and stability, the isentropic \emph{speed of sound}, $c_s$, must be real and not exceed the speed of light:
$$c_s^2 = \left( \frac{\partial p}{\partial \mu} \right)_S \qquad \text{must satisfy} \qquad 0 \le c_s^2 \le 1$$

\subsection{Decomposition of the Electromagnetic and Weyl Tensors}

To build intuition for the 1+3 decomposition of gravity, we first review the well-known case of electromagnetism, which serves as a powerful analogy.

\subsubsection{Electromagnetic Fields}

The electromagnetic field strength tensor, $F_{ab}$, is an anti-symmetric 2-tensor. An observer with 4-velocity $u^a$ will naturally measure this field as distinct electric and magnetic components. These are defined as the spatial vectors:
\begin{enumerate}
\item \underline{\it  Electric Field:} The force experienced by a unit charge at rest with the observer.
$$E_a = F_{ab} u^b \quad \implies \quad E_a u^a = 0$$
\item \underline{\it  Magnetic Field:} The spatial dual of the remaining components of $F_{ab}$.
$$H_a = \frac{1}{2} \eta_{abc} F^{bc} \quad \implies \quad H_a u^a = 0$$
\end{enumerate}

Here, $\eta_{abc}$ is the spatial volume element, ensuring both $E_a$ and $H_a$ are purely spatial vectors living in the observer's rest frame.

\subsubsection{Weyl Tensor}
\label{sec:WeylTensor}

In direct analogy, the Weyl conformal curvature tensor, $C_{abcd}$, which represents the free gravitational field (tidal forces and gravitational waves), can be decomposed into its \emph{electric} and \emph{magnetic} parts~\cite{1989-Ellis.Bruni-PRD}. Because gravity is a spin-2 field, these components are not vectors, but spatial, symmetric, and trace-free (SSTF) tensors.
\begin{enumerate}
\item \underline{\it  Electric Weyl Tensor:} Represents tidal forces—the stretching and squeezing of bodies.
\begin{equation}
 E_{ab} = C_{acbd} u^c u^d
 \label{eq:ElectricWeyl}
\end{equation}
\item \underline{\it  Magnetic Weyl Tensor:} Represents frame-dragging effects and gravitational radiation.
\begin{equation}
H_{ab} = \frac{1}{2} \eta_{ade} C^{de}{}_{bc} u^c
\label{eq:MagneticWeyl}
\end{equation}
\end{enumerate}
Both tensors are purely spatial ($E_{ab}u^b = 0$, $H_{ab}u^b = 0$) and SSTF ($E_{ab} = E_{\langle ab \rangle}$, $H_{ab} = H_{\langle ab \rangle}$).

\subsection{The 1+3 gravity field equations}
\label{sec:1+3Decom} 

The fundamental equations of GR can now be rewritten in this 1+3 language. The Ricci and Bianchi identities, when decomposed, provide a complete set of evolution and constraint equations for the matter and gravitational fields.

\noindent \underline{\it  Kinematical Evolution (from Ricci Identities)} The Ricci identities ($2 \nabla_{[a} \nabla_{b]} u_c = R_{abc}{}^d u_d$) govern the evolution of the fluid's kinematical properties. They split into two sets:
\begin{enumerate}
    \item \underline{\it  Propagation Equations (Evolution in Time)}
\begin{enumerate}
\item \emph{Raychaudhuri Equation (Expansion Propagation):}
    This describes the acceleration of the volume expansion.
\begin{equation}
    \dot{\Theta} + \frac{1}{3}\Theta^2 = \widetilde{\nabla}_a \dot{u}^a - \dot{u}_a \dot{u}^a + 2\sigma^2 - 2\omega^2 - \frac{1}{2}(\mu + 3p) + \Lambda
\label{eq:Raychaudhuri}
\end{equation}
    This is the fundamental equation of gravitational attraction in GR. It shows that matter with density $\mu$ and pressure $p$ causes expansion to decelerate, while shear and vorticity can resist collapse.

\item \emph{Shear Propagation:}
    This describes how shear evolves, driven by gravitational tides and anisotropic pressure.
\begin{equation}
    \dot{\sigma}_{\langle ab \rangle} = -\frac{2}{3}\Theta \sigma_{ab} - \sigma_{c\langle a}\sigma_{b\rangle}{}^c - \omega_{\langle a}\omega_{b\rangle} + \widetilde{\nabla}_{\langle a}\dot{u}_{b\rangle} + \dot{u}_{\langle a}\dot{u}_{b\rangle} - \left( E_{ab} - \frac{1}{2}\pi_{ab} \right)
\label{eq:Shear}
\end{equation}
    This key equation shows that the electric part of the Weyl tensor ($E_{ab}$) and anisotropic stress ($\pi_{ab}$) act as sources for shear.
    
\item \emph{Vorticity Propagation:}
    This describes how vorticity evolves.
\begin{equation}
    \dot{\omega}_{\langle a \rangle} = -\frac{2}{3}\Theta \omega_a + \sigma_{ab} \omega^b + \frac{1}{2} \eta_{abc} \widetilde{\nabla}^b \dot{u}^c
\label{eq:Vorticity}    
\end{equation}
    If the acceleration is curl-free (e.g., derived from a potential), the last term vanishes, showing that vorticity simply decays with the universe's expansion. Unlike shear and expansion, if there is no vorticity initially for a geodesic trajectory, then there is no vorticity evolution.
\end{enumerate}

\item \underline{\it  Constraint Equations (Conditions on a Spatial Slice)}
\begin{enumerate}
\item \emph{Momentum-Density Constraint:}
    This relates the spatial gradient of expansion and vorticity to the energy flux.
    $$
    \widetilde{\nabla}^b \sigma_{ab} - \frac{2}{3}\widetilde{\nabla}_a \Theta + \eta_{abc} ( \widetilde{\nabla}^b \omega^c + 2\dot{u}^b \omega^c ) = -q_a
    $$

\item \emph{Vorticity Constraint:}
    The divergence of vorticity is constrained by the acceleration.
    $$
    \widetilde{\nabla}_a \omega^a = \dot{u}_a \omega^a
    $$

\item \emph{Magnetic Weyl Constraint:}
    The magnetic part of the Weyl tensor is generated by the curl of the shear and vorticity gradients.
    $$
    H_{ab} = (\text{curl}~\sigma)_{ab} - \widetilde{\nabla}_{\langle a}\omega_{b\rangle} - 2 \dot{u}_{\langle a} \omega_{b \rangle} \quad \text{where} \quad (\text{curl}~\sigma)_{ab} = \eta_{cd(a} \widetilde{\nabla}^c \sigma_{b)}{}^d
    $$
\end{enumerate}
\end{enumerate}

\subsection{Gravitational Field Dynamics (from Bianchi Identities)}
\label{sec:1+3-FieldDynamics}

The final set of dynamical equations arises from the contracted Bianchi identities ($\nabla_\ell C_{abcd} +  \nabla_n C_{a b \ell m}+ 
\nabla_m C_{a b n \ell}=0$). These identities govern the evolution of the free gravitational field, represented by the electric ($E_{ab}$) and magnetic ($H_{ab}$) parts of the Weyl tensor. The resulting equations bear a striking resemblance to Maxwell's equations, providing a complete description of how gravitational fields are generated by matter and how they propagate through spacetime. The decomposition yields two propagation equations and two constraint equations.

\subsubsection{Propagation Equations: The Gravitational "Induction Laws"}

These two equations describe how the electric and magnetic gravitational fields evolve and induce one another, forming the basis for gravitational waves.
\begin{enumerate}
\item \underline{\it  Evolution of the Electric Weyl Tensor (Ampere-Maxwell Analogue):} 

This equation shows that the electric Weyl field is generated by the curl of the magnetic Weyl field, as well as by matter sources like anisotropic stress ($\pi_{ab}$) and heat flux ($q_a$).
    \begin{align}
    \dot{E}_{\langle ab \rangle} - (\text{curl}~H)_{ab} = &-\frac{1}{2} \dot{\pi}_{\langle ab \rangle} - \frac{1}{2} \widetilde{\nabla}_{\langle a} q_{b \rangle} -\frac{1}{2} (\mu + p) \sigma_{ab} - \Theta \left( E_{ab} + \frac{1}{6} \pi_{ab} \right) \nonumber \\
    & + 3 \sigma_{\langle a}{}^c \left( E_{b \rangle c} - \frac{1}{6} \pi_{b \rangle c} \right) - \dot{u}_{\langle a} q_{b \rangle} \nonumber \\
    & + \eta^{cd}{}_{\langle a} \left[ 2 \dot{u}_c H_{b \rangle d} + \omega_c \left( E_{b \rangle d} + \frac{1}{2} \pi_{b \rangle d} \right) \right]
    \end{align}

\item \underline{\it  Evolution of the Magnetic Weyl Tensor (Faraday's Law Analogue):}

This equation shows that the magnetic Weyl field is generated by the curl of the electric Weyl field and the curl of anisotropic stress.
    \begin{align}
    \dot{H}_{\langle ab \rangle} + (\text{curl}~E)_{ab} = &~ \frac{1}{2} (\text{curl}~\pi)_{ab} -\Theta H_{ab} + 3 \sigma_{\langle a}{}^c H_{b \rangle c} + \frac{3}{2} \omega_{\langle a} q_{b \rangle} \nonumber \\
    & - \eta^{cd}{}_{\langle a} \left[ 2 \dot{u}_c E_{b \rangle d} - \frac{1}{2} \sigma_{b \rangle c} q_d - \omega_c H_{b \rangle d} \right]
    \end{align}
\end{enumerate}

The `curl` operation for any spatial, symmetric, trace-free tensor $T_{ab}$ is defined as $(\text{curl}~T)_{ab} = \eta_{cd\langle a} \widetilde{\nabla}^c T_{b \rangle}{}^d$.

These two coupled equations are the heart of gravitational wave dynamics. By taking a time derivative of one and substituting the other, one can derive a second-order hyperbolic wave equation for both $E_{ab}$ and $H_{ab}$. This formally demonstrates that gravitational radiation propagates through spacetime as interacting ripples of tidal ($E_{ab}$) and frame-dragging ($H_{ab}$) fields.

\subsubsection{Constraint Equations: The Gravitational ``Gauss's Laws"}

These two equations do not involve time derivatives. Instead, they constrain the gravitational field on any given spatial hypersurface, linking its structure directly to the matter distribution.
\begin{enumerate}
\item \underline{\it  Electric Weyl Constraint (Gravitational Gauss's Law):}
    The divergence of the electric Weyl tensor is sourced by the spatial gradients of the energy density. This is the relativistic analogue of Newton's Poisson equation, enabling tidal action at a distance.
    \begin{equation}
    \widetilde{\nabla}_b E^{ab} = \frac{1}{3}\widetilde{\nabla}^a\mu - \widetilde{\nabla}_b\left(\frac{1}{2}\pi^{ab}\right) - \frac{1}{3}\Theta q^a + \frac{1}{2}\sigma^a{}_b q^b + 3\omega_b H^{ab} + \eta^{abc}\left[\sigma_{bd}H_c{}^d - \frac{3}{2}\omega_b q_c\right]
    \end{equation}

\item \underline{\it  Magnetic Weyl Constraint:}
    The divergence of the magnetic Weyl tensor is sourced primarily by the vorticity and heat flux of the matter fluid.
    \begin{equation}
    \widetilde{\nabla}_b H^{ab} = -(\mu+p)\,\omega^a - 3\,\omega_b\left(E^{ab}-\tfrac{1}{6}\pi^{ab}\right) - \eta^{abc}\left[\tfrac{1}{2}\widetilde{\nabla}_b q_c+\sigma_{bd}\left(E_c{}^d+\tfrac{1}{2}\pi_c{}^d\right)\right]
    \end{equation}
\end{enumerate}

These constraints link the gravitational field to the nature of cosmological perturbations. Inhomogeneities in energy density \emph{ (scalar modes)} source the divergence of $E_{ab}$, while rotational fluid flows (\emph{vector modes}) are the primary sources for the divergence of $H_{ab}$.

\subsection{Matter Conservation Equations}
\label{Energy}

The basis of fluid dynamics in GR is the covariant conservation of the energy-momentum tensor, which states that there are no local sources or sinks of energy-momentum:
\begin{equation}
   \nabla_b T^{ab} = 0 
\end{equation}
In the 1+3 formalism, this single 4-dimensional equation is projected into two separate equations with clear physical interpretations. Projecting parallel to the observer's 4-velocity ($u_a \nabla_b T^{ab}=0$) yields the law of energy conservation, while projecting orthogonally into the rest space ($h_{ac} \nabla_b T^{cb}=0$) yields the law of momentum conservation.

\subsubsection{Energy Conservation}

This equation is the relativistic form of the first law of thermodynamics, describing how the energy density in a comoving volume changes.
\begin{equation}
\dot{\mu} + (\mu + p)\Theta = - \widetilde{\nabla}_a q^a - 2\dot{u}_a q^a - \sigma_{ab}\pi^{ab}
\end{equation}
Each term has a distinct physical interpretation:
\begin{enumerate}
    \item  $\dot{\mu}$: The rate of change of internal energy density as measured by the observer.
    \item $(\mu + p)\Theta$: This term describes the work done by the fluid. It combines the dilution of energy density due to volume expansion ($\mu\Theta$) and the thermodynamic work done by pressure as the volume changes ($p\Theta$).
    \item $-\widetilde{\nabla}_a q^a$: This is the net energy flowing into the local volume due to heat flux (the divergence of $q^a$).
    \item The remaining terms on the right-hand side are dissipative effects, representing work done by non-inertial forces on the heat flux and energy loss due to viscous friction ($\sigma_{ab}\pi^{ab}$).
\end{enumerate}

\subsubsection{Momentum Conservation (Relativistic Euler Equation)}

This equation is the relativistic and covariant generalization of the Navier-Stokes (or Euler, for a perfect fluid) equation. It describes the forces acting on a fluid element.
\begin{equation}
(\mu + p)\dot{u}_a + \widetilde{\nabla}_a p + \widetilde{\nabla}_b \pi_{ab} + \dot{q}_{\langle a \rangle} + \frac{4}{3}\Theta q_a + \sigma_{ab} q^b + \eta_{abc}\omega^b q^c = 0
\end{equation}
To understand this equation, it's helpful to first consider a \emph{perfect fluid} ($\pi_{ab}=0, q_a=0$), for which it simplifies  to:
$$(\mu + p)\dot{u}_a = - \widetilde{\nabla}_a p$$
This is a direct relativistic analogue of Euler's equation ($\rho \mathbf{a} = -\nabla p$). It states that a pressure gradient is the only force that can accelerate the fluid. The term $(\mu+p)$ plays the role of the \emph{inertial mass density}.

The additional terms in the full equation are corrections for:
\emph{Viscous forces} from anisotropic stress ($\widetilde{\nabla}_b \pi_{ab}$) and {\emph Momentum associated with heat flow} ($q_a$) and its interaction with the expanding, shearing, and rotating background.

\section{The Null Formalism: A $1+2+1$ Perspective}
\label{sec:NullFormalism}

While the 1+3 formalism is perfect for describing physics from the perspective of massive, timelike observers, it fundamentally breaks down on null surfaces like black hole event horizons, where the normal vector is lightlike. To analyze these critical regions, we must adapt our framework. This leads to the \emph{null formalism}, which splits spacetime not into time and space, but into two null directions and a 2-dimensional transverse \emph{screen space.}

\subsection{The Null Dyad and Screen Space}

We select a congruence of null geodesics with tangent vector field $k^a$ ($k^a k_a = 0$). To define the 2D space orthogonal to this direction of propagation, we introduce an auxiliary null vector $l^a$ ($l^a l_a = 0$), chosen such that it is normalized against $k^a$:
$$k^a l_a = -1$$
Together, the pair $(k^a, l^a)$ forms a \emph{null dyad}. This allows us to define the \emph{screen space projection tensor}, $N_{ab}$, which projects vectors onto the 2D space orthogonal to both $k^a$ and $l^a$:$$N_{ab} = g_{ab} + k_a l_b + l_a k_b$$
This tensor satisfies $N_{ab}k^b = 0$ and $N_{ab}l^b = 0$.

\subsection{Kinematics of Null Congruences}

Just as we did for timelike congruences, we can describe the geometrical properties of a null flow by decomposing the derivative of its tangent vector. Projecting $\nabla_b k_a$ onto the screen space gives:
\begin{equation}
N_a{}^c N_b{}^d \nabla_c k_d = \frac{1}{2} \theta N_{ab} + \sigma_{ab} + \omega_{ab}  
\end{equation}
Here again, each term has a distinct physical interpretation:
\begin{enumerate}
\item \underline{\it  Expansion ($\theta = N^{ab} \nabla_a k_b$):} Measures the expansion or contraction of the area of a beam of light rays.
\item \underline{\it  Shear ($\sigma_{ab}$):} Measures the distortion of the shape of the beam's cross-section.
\item \underline{\it  Twist ($\omega_{ab}$):} Measures the rotation of the light rays around each other.
\end{enumerate}

The evolution of these kinematical quantities along the null geodesics (parameterized by an affine parameter $\lambda$) is governed by the \emph{null Ricci equations}:
\begin{enumerate}
\item \underline{\it  Null Raychaudhuri Equation (Expansion Evolution):}
\begin{equation}
k^a \nabla_a \theta \equiv \frac{d\theta}{d\lambda} = -\frac{1}{2} \theta^2 - \sigma_{ab} \sigma^{ab} + \omega_{ab} \omega^{ab} - R_{ab} k^a k^b
\end{equation}
This shows that focusing of light rays ($d\theta/d\lambda < 0$) is guaranteed by the presence of matter ($R_{ab}k^a k^b > 0$) and shear (when the twist is absent).

\item \underline{\it  Shear and Twist Evolution:}
\begin{eqnarray}
\frac{d\sigma_{ab}}{d\lambda} &\equiv&  k^c \nabla_c \sigma_{ab} = -\theta \sigma_{ab} - C_{acbd} k^c k^d + \dots \\ 
\frac{d\omega_{ab}}{d\lambda} &\equiv & k^c \nabla_c \omega_{ab} = -\theta \omega_{ab} - 2 \sigma_{[a}{}^c \omega_{b]c}
\end{eqnarray}
The shear is sourced by the electric part of the Weyl tensor projected along the null rays ($C_{acbd} k^c k^d$), while the twist evolution depends on shear. For clarity, we have omitted terms quadratic in shear and twist.
\end{enumerate}

\begin{table}
\centering
\caption{Key differences between time-like and null kinematics}
\label{tab:Table}
\begin{tblr}{
  width = \linewidth,
  colspec = {Q[192]Q[330]Q[427]},
  hlines,
  vlines,
}
{\bf Feature} & {\bf Time-like congruence} ($u^a$)  & {\bf Null congruence} ($k^a$)   \\
Physical System  & The worldlines of massive observers or fluid particles.          & The paths of light rays (null geodesics).   \\
Spacetime Split  & 1+3 Decomposition: Spacetime is split into a unique time direction and a 3D spatial rest-space. & 1+2+1 Decomposition: Spacetime is split into a null propagation direction, an auxiliary null direction, and a 2D transverse screen space.        \\
Expansion ($\Theta$ vs. $\theta)$ & Measures the change in the 3D volume of a small ball of observers.  & Measures the change in the 2D area of the cross-section of a beam of light. \\
Shear ($\sigma_{a b}$)   & Describes the distortion of a 3D ball of observers into an ellipsoid.  & Describes the distortion of a 2D shape (e.g., a circle into an ellipse) in the light beam's cross-section. \\
Rotation ($\omega_{a b})$ & Vorticity: Describes the swirling or rigid rotation of the 3D fluid element around an axis.  & Twist: Describes the rotation of the 2D screen space itself as the beam propagates. It measures if the light rays are twisting around each other. 
\end{tblr}
\end{table}

\subsection{Timelike vs. Null Kinematics}

While the kinematic quantities derived from the null vector $k^a$ and timelike vector $u^a$ share similar names and arise from a similar mathematical procedure (decomposing the covariant derivative of the congruence's tangent vector), they describe the behavior of fundamentally different physical systems in different dimensions.

It may be easy to understand the difference through an analogy: The timelike ($u^a$) kinematics describe the flow of a fluid, like a blob of fluid~\cite{2012-ellis_maartens_maccallum-Book}. However, the null ($k^a$) kinematics describe the propagation of light, like the beam from a projector. Table \eqref{tab:Table} gives key differences between timelike and null kinematics. 

\subsection{Gravitational Fields in the Null Formalism}

The Bianchi identities provide the evolution equations for the Weyl tensor components, defined with respect to the null dyad. These equations, referred as \emph{null Bianchi identities}, govern how gravitational radiation propagates along null rays.
\begin{enumerate}
\item \underline{\it  Propagation Equations:}
\begin{eqnarray}
& & \frac{dE_{ab}}{d\lambda} - \text{curl } H_{ab} = -\frac{1}{2}(\mu + p) \sigma_{ab} - \theta E_{ab} + \dots \\ 
& & \frac{dH_{ab}}{d\lambda} + \text{curl } E_{ab} = -\theta H_{ab} + \dots
\end{eqnarray}

\item \underline{\it  Constraint Equations:}
\begin{eqnarray}
& & \nabla^b E_{ab} = \frac{1}{3}\nabla_a \mu - 3\omega^b H_{ab} + \dots \\ 
& & \nabla^b H_{ab} = -(\mu+p)\omega_a + \dots     
\end{eqnarray}
Here $E_{ab}$ and $H_{ab}$ are the electric and magnetic parts of the Weyl tensor defined relative to the null dyad, and we have shown only the leading-order terms for brevity. These equations form the basis for studying gravitational waves and near-horizon physics~\cite{Chakraborty:2024ars,Bhattacharya:2022ncn}.
\end{enumerate}

\subsection{Maxwell's Equations: A Structural Template}

To build intuition, we present the fully covariant decomposition of Maxwell's equations in both $(1 + 3)$ and $1+2+1$ formalisms~\cite{1996-vanElst.Ellis-CQG,Jana:2023kyq}. Their structure provides a clear and simple template for their more complex gravitational counterparts. The fundamental equations are:
\begin{equation}
\nabla_b F^{ab} = j^a, \qquad \nabla_{[a} F_{bc]} = 0
\end{equation}

\subsubsection{Timelike (1+3) Decomposition}

Relative to a timelike observer $u^a$, the equations split as:
\begin{enumerate}
    \item \underline{\it  Propagation:}
\begin{eqnarray}
& & \dot{E}_{\langle a \rangle} - \eta_{abc} \widetilde{\nabla}^b H^c = - j_{\langle a \rangle} - \Theta E_a + \dots \\   
& & \dot{H}_{\langle a \rangle} + \eta_{abc} \widetilde{\nabla}^b E^c = - \Theta H_a + \dots    
\end{eqnarray}
\item \underline{\it  Constraints:}
\begin{eqnarray}
& & \widetilde{\nabla}_a E^a = \rho_e + 2 \omega_a H^a \\   
& & \widetilde{\nabla}_a H^a = - 2 \omega_a E^a
\end{eqnarray}

\end{enumerate}
These equations show how spacetime kinematics (expansion $\Theta$, vorticity $\omega_a$) act as effective sources for the electromagnetic field.

\subsubsection{Null (1+2+1) Decomposition}

Relative to a null congruence $k^a$, the equations take a nearly identical form:
\begin{enumerate}
    \item \underline{\it  Propagation:}
\begin{eqnarray}
& &  \frac{dE_a}{d\lambda} - \eta_{abc} \nabla^b H^c = - j_a - \frac{1}{2}\theta E_a + \dots \\ 
& & \frac{dH_a}{d\lambda} + \eta_{abc} \nabla^b E^c = - \frac{1}{2}\theta H_a + \dots
\end{eqnarray}
\item \underline{\it  Constraints:}
\begin{eqnarray}
& &  \nabla_a E^a = \rho_e + 2 \omega_a H^a \\
& & \nabla_a H^a = - 2 \omega_a E^a
\end{eqnarray}
\end{enumerate}
The structure is preserved, with the null kinematical quantities ($\theta, \omega_a$) replacing their timelike analogues. This remarkable structural similarity between the timelike, null, electromagnetic, and gravitational systems is a deep feature that we will exploit in our analysis of the Carrollian limit.

\section{Galilean and Carrollian Limits of Spacetime}
\label{sec:GalileanCarrollian}

To build our formalism, we begin with the standard Lorentz transformations between two inertial frames (unprimed and primed) moving with a relative velocity $\mathbf{v}$:
\begin{equation}
\mathbf{x}' = \mathbf{x} - \gamma \frac{\mathbf{v}}{c} ct + (\gamma -1) \frac{\mathbf{v} (\mathbf{v} \cdot \mathbf{x})}{v^2}, \quad ct' = \gamma \left[ ct - \frac{\mathbf{v}}{c} \cdot \mathbf{x} \right], 
\label{def:Lorenz-Trans}
\end{equation}
where $\gamma = (1 - v^2/c^2)^{-1/2}$.
\subsection{The Structure of Spacetime Limits: Dynamical vs. Kinematic}

The Lorentz transformations admit several distinct limits, which can be broadly categorized into two types: \emph{dynamical limits}, where the constant $c$ is treated as a variable, and \emph{kinematic limits}, where $c$ is fixed and the geometry is probed in specific spacetime regions.

\subsubsection{The Dynamical Limits: $c \to \infty$ and $c \to 0$}

The most direct way to deform the structure of spacetime is by altering the value of the speed of light itself. This gives rise to the two foundational limits:
\begin{enumerate}
\item \underline{\it  The Galilean Limit ($c \to \infty$):} This is the \emph{instantaneous limit}. By letting $c \to \infty$ in the Lorentz transformations, the time transformation becomes $t' = t$, establishing an absolute, universal time. The spatial transformation becomes $\mathbf{x}' = \mathbf{x} - \mathbf{v}t$. In this limit, the light cone opens up to fill the entire spacetime, meaning interactions are instantaneous. This is the symmetry group of Newtonian physics. 

\item \underline{\it  The Carrollian Limit ($c \to 0$):} This is the \emph{ultra-local limit}. As $c \to 0$, the light cone collapses to the time axis. Causal contact between spatially separated points becomes impossible. The transformations become $\mathbf{x}' = \mathbf{x}$ and $t' = t - (\mathbf{v} \cdot \mathbf{x})/c^2$, establishing an absolute sense of space but making dynamics local to each worldline. 
\end{enumerate}

\subsubsection{The Kinematic Limits: Low Velocity Regimes}

A more subtle approach is to keep $c$ fixed and consider the low-velocity limit ($|\mathbf{v}|/c \ll 1$) in different regions of spacetime. This reveals that the Galilean and Carrollian structures also emerge as descriptions of specific physical regimes.
\begin{enumerate}
\item \underline{\it  The Ultra-timelike Limit ($ct \gg |\mathbf{x}|$):} In this region, where events are separated by large temporal and small spatial intervals, the low-velocity limit of the Lorentz transformations once again yields the \emph{Galilean transformations}. Even though the underlying physics is relativistic, for observers in this regime, spacetime \emph{appears} Galilean. In this case, $\gamma \to 1$, and the Lorentz transformations 
\eqref{def:Lorenz-Trans} immediately reduce to the familiar Galilean form:
\begin{equation}
\label{def:Galilean-Trans}
\mathbf{x}' = \mathbf{x} - \frac{\mathbf{v}}{c}{ct}, \quad t' = t.
\end{equation} 
For a boost in the x-direction, this yields a transformation matrix:
\begin{equation}
\label{matrix:Galilean-Trans}
 \Lambda_g = \begin{pmatrix}
     1&0&0&0\\
     -\frac{v}{c}&1&0&0\\
     0&0&1&0\\
     0&0&0&1
 \end{pmatrix} = I+A   
\end{equation}
Here \(A\) matrix has only one non-zero non-diagonal component (\(A^1_0 =-\frac{v}{c}\)). It can be seen that \(A^2 = 0\), and thus \(\Lambda_g^{-1} = I-A\). 
Vectors transform as $x'^a = \Lambda_g^a{}_b x^b$, while dual vectors transform with the inverse, $\Lambda_g^{-1}$.
\item \underline{\it  The Ultra-spacelike Limit ($ct \ll |\mathbf{x}|$):} In this region, where events are separated by large spatial and small temporal intervals, the low-velocity limit yields the \emph{Carrollian transformations}:
\begin{equation}
\label{def:Carrollian-Trans}
\mathbf{x}' = \mathbf{x}, \quad t' = t - \frac{1}{c^2} \mathbf{v} \cdot \mathbf{x}.
\end{equation}
For a boost in the x-direction, the transformation matrix takes the form:
\begin{equation}
\label{matrix:Carrollian-Trans}
 \Lambda_c = \begin{pmatrix}
     1&-\frac{v}{c}&0&0\\
     0&1&0&0\\
     0&0&1&0\\
     0&0&0&1
 \end{pmatrix} = I+B   
\end{equation}
Like $A$, \(B\) matrix has only one non-zero component (\(B^0_1 =-\frac{v}{c}\)). It can be seen that \(B^2 = 0\), and thus \(\Lambda_c^{-1} = I-B\). Vectors are transformed using \(\Lambda_c\), while dual vectors are transformed using its inverse - \(\Lambda_c^{-1}\).
In particular, causal relations between events seem to be impossible.These explicit transformation matrices form the foundation for systematically applying the Carrollian limit to the $(1+3)-$ decomposed field equations in the following sections.
\end{enumerate}

This reveals a key insight: the \emph{Galilean group} can be seen as both a fundamental structure (from $c \to \infty$) and as an effective description in the ultra-timelike, low-velocity regime.

As we will show, this branching of limits has important consequences for field theories. The attempt to formulate Maxwell's equations within the Galilean spacetime framework reveals that a single, unified theory is not possible. This forces a split into two distinct, consistent non-relativistic theories—the electric limit and the magnetic limit—each defined by an additional set of physical \emph{dominance conditions} that specify which fields and sources are primary.
Our work is predicated on the idea that a similar structure exists for gravity within the \emph{Carrollian limit}, which we will now explore.

\section{Galilean Limits of Maxwell's Electromagnetism}
\label{sec:Galilean-EM}

In this section, we demonstrate that a single, unified Galilean limit of Maxwell's theory does not exist. Instead, the low-velocity limit ($v \ll c$) splits into two distinct, physically relevant frameworks --- the \emph{Electric Limit} and the \emph{Magnetic Limit}. We will derive the transformation laws for each and find the corresponding set of covariant equations that remain invariant under those laws~\cite{Heras_2010}. We will work in units where $c=1$. 

Our starting point is the full set of 1+3 covariant Maxwell's equations on a generic curved background, sourced by a charge density $\rho_e = -j^a u_a$ and a spatial current $I^a = h^a_b j^b$.
\begin{subequations}
\label{eq:Maxwell}
\begin{eqnarray} 
\widetilde{\nabla}_a E^a = \rho_e & & \qquad \mbox{(Gauss law for Electric field)}  \\
\widetilde{\nabla}_a H^a = 0 & & \qquad \mbox{(Gauss law for Magnetism)} \\
\dot{H}^{\langle a \rangle} + \frac{2}{3}\theta H^a = -\eta^{abc}\widetilde{\nabla}_b E_c & & \qquad
\mbox{(Faraday's Law)} \\
\dot{E}^{\langle a \rangle} + \frac{2}{3}\theta E^a = \eta^{abc}\widetilde{\nabla}_b H_c - I^a & & \qquad
\mbox{(Ampere-Maxwell Law)}
\end{eqnarray} 
\end{subequations}
These equations are fully Lorentz covariant. The fully relativistic transformation of the electric, magnetic fields, charge and current densities are:
\begin{eqnarray}
\label{ELorentz}
E'^i &=& \gamma \left( E^i + \frac{c_u^2}{c} \eta^{ijk} v_j H_k \right) 
      - \frac{\gamma - 1}{v^2} \, v^i ( v^m E_m )  \\
H'^i &=& \gamma \left( H^i - \frac{c}{c_u^2} \eta^{ijk} v_j E_k \right) 
      - \frac{\gamma - 1}{v^2} \, v^i ( v^m H_m ) \\
c \rho_e' &=& \gamma \left( c\rho_e - v^i I_i \right) \\
I'^i &=& \gamma \left( I^i - \frac{v^i}{c} c\rho_e \right) 
      + \frac{\gamma - 1}{v^2} \, v^i ( v^m I_m ). \label{ILorentz}
\end{eqnarray}
where $(i,j,k)$ are the  spatial components and the temporal components do not change (considered to be zero).

{We note that the above 1+3 covariant Maxwell's equations are presented in a general form, valid for the curved and expanding FLRW background. A natural question is how these equations relate to the more familiar flat-space limits, such as those discussed in Ref. \cite{Duval:2014uoa}. Our equations reduce to the flat-space case by \emph{specializing the background}. This is achieved by setting all background kinematic quantities to zero ($\theta = 0$, $\sigma_{ab} = 0$, $\omega_{ab}=0$, $\dot{u}_a=0$) and choosing a comoving inertial frame where $u^a = \delta^a_0$. In this limit, the covariant derivatives become standard partial derivatives: $\dot{T} \to \partial_t T$, $\widetilde{\nabla}_a E^a \to \nabla \cdot \mathbf{E}$, and $\eta^{abc}\widetilde{\nabla}_b E_c \to \nabla \times \mathbf{E}$. The Galilean and Carrollian limits we derive in the following sections will, under this specialization, precisely reduce to their well-known flat-space counterparts. Our work thus provides the covariant generalization of these limits to curved spacetimes.}
We will now test their invariance under the two different Galilean limits.

\subsection{The Electric Limit}

As mentioned earlier, the Electric Limit describes situations where electric effects are dominant. This limit is defined by the condition:
\begin{equation}
|E| \gg |H| \qquad \text{and} \qquad |\rho_e| \gg |I|   
\end{equation}
%
Applying these conditions to the full Lorentz transformations for the fields and sources in the low-velocity limit ($v \ll 1, \gamma \to 1$) yields the \emph{electric transformation rules}:
\begin{align}
E'^a = E^a \, ;  \quad \rho'_e = \rho_e \, ; & \quad 
H'^a = H^a - \eta^{abc}v_b E_c \, ;   \quad 
I'^a = I^a - v^a \rho_e \, .
\end{align}
%

We now check if the full set of Maxwell's equations remains invariant under these transformations. A direct calculation shows that Gauss's Law and the Ampere-Maxwell Law are invariant. However, Faraday's Law and Gauss's Law for Magnetism are \emph{not}. For example, transforming Gauss's Law for Magnetism gives:
\begin{equation}
\widetilde{\nabla}_a H'^a = \widetilde{\nabla}_a (H^a - \eta^{abc}v_b E_c) = \widetilde{\nabla}_a H^a - v_b (\eta^{bac}\widetilde{\nabla}_a E_c)   
\end{equation}
For the law $\widetilde{\nabla}_a H'^a = 0$ to hold, we require the original $\widetilde{\nabla}_a H^a = 0$ and the non-invariant term to vanish.
%
%
To construct a consistent, invariant theory, we must demand that the terms breaking the invariance are zero. In the example above, this forces a new physical constraint:
\begin{equation}
\eta^{bac}\widetilde{\nabla}_a E_c = \text{curl}(E)^b = 0
\end{equation}
Applying this systematically, we find that Faraday's law must be replaced by this new constraint. The result is the \emph{Electric Limit of Maxwell's Equations}:
\begin{eqnarray}
\widetilde{\nabla}_a E^a = \rho_e & & \qquad \mbox{(Gauss's Law)}   \\
\eta^{abc}\widetilde{\nabla}_b E_c = 0 & & \qquad 
\mbox{(Constraint on E)} \\
\widetilde{\nabla}_a H^a = 0 & & \qquad \mbox{(Gauss's Law for Magnetism)} \\
\dot{E}^{\langle a \rangle} + \frac{2}{3}\theta E^a = \eta^{abc}\widetilde{\nabla}_b H_c - I^a & & \qquad
\mbox{(Ampere-Maxwell Law)}
\end{eqnarray}
The invariance of these equations can be seen in Eqs. \ref{eq:GalileanElec-EM-Invariance}.
This theory describes \emph{electro-quasistatics}, where magnetic induction is negligible and governs capacitive effects. Note that in this limit the electric fields is purely irrotational.

\subsection{The Magnetic Limit}

As described early, the Magnetic Limit describes situations where magnetic effects and currents are dominant. This limit is defined by the conditions:
$$|H| \gg |E| \qquad \text{and} \qquad |I| \gg |\rho_e|$$ These lead to the \emph{magnetic transformation rules}:
%
\begin{align}
E'^a = E^a + \eta^{abc}v_b H_c; \quad \rho'_e = \rho_e - v_a I^a; & \quad H'^a = H^a; \quad I'^a = I^a
\end{align}

Following the same procedure, we find that under these transformations, Faraday's Law and Gauss's Law for magnetism are invariant, but the other two laws are not. To restore invariance, we must impose new constraints. The most significant change is that the Ampere-Maxwell law must be truncated, effectively dropping the displacement current term ($\dot{E}$).

The resulting \emph{Magnetic Limit of Maxwell's Equations} is:
\begin{eqnarray}
\widetilde{\nabla}_a E^a = \rho_e  & & \qquad \mbox{(Gauss's Law)}   \\
\dot{H}^{\langle a \rangle} + \frac{2}{3}\theta H^a = -\eta^{abc}\widetilde{\nabla}_b E_c & & \qquad
\mbox{(Faraday's Law)} \\
\widetilde{\nabla}_a H^a = 0 & & \qquad \mbox{(Gauss's Law for Magnetism)} \\
\eta^{abc}\widetilde{\nabla}_b H_c = I^a  & & \qquad
\mbox{(Truncated Ampere's Law)}
\end{eqnarray}
The invariance of these equations can be seen in Eqs. \ref{eq:GalileanMag-EM-Invariance}. Note that in the last equation, Curl of $H$ is sourced by current. This theory describes \emph{magneto-quasistatics} and corresponds to ``electrodynamics before Maxwell," where the displacement current had not yet been discovered. It governs inductive effects and is the foundation for \emph{magnetohydrodynamics}~\cite{Chandrasekhar1961MHD}.

\section{The Instantaneous Limit and its Relation to the Electric Limit}
\label{sec:Instat-EM}

In the previous section we showed how the low-velocity limit ($v \ll c$) of Maxwell's theory splits into two distinct kinematic regimes—the Electric and Magnetic limits. We now consider a different approach: the \emph{dynamical limit} where the speed of light itself is taken to infinity ($c \to \infty$). This is known as the \emph{instantaneous limit}, as it describes a hypothetical world where all interactions propagate instantly.

Applying the $c \to \infty$ limit to the fully relativistic Lorentz transformations for the fields (Eqs. \ref{ELorentz} - \ref{ILorentz}) yields the following transformation rules:
\begin{align}
E'^a = E^a; \quad 
H'^a = H^a - \eta^{abc} v_b E_c; & \quad 
\rho'_e = \rho_e; \quad 
I'^a = I^a - v^a \rho_e
\end{align}
Interestingly, these are formally identical to the transformation laws we derived for the \emph{Electric Limit}. Consequently, the set of Maxwell's equations that remains invariant under these transformations is also identical to the Electric Limit equations:
\begin{eqnarray}
\widetilde{\nabla}_a E^a = \rho_e & & \qquad \mbox{(Gauss's Law)}   \\
\eta^{abc}\widetilde{\nabla}_b E_c = 0 & & \qquad
\mbox{(Constraint on E)} \\
\widetilde{\nabla}_a H^a = 0 & & \qquad \mbox{(Gauss's Law for Magnetism)} \\
\dot{E}^{\langle a \rangle} + \frac{2}{3}\theta E^a = \eta^{abc}\widetilde{\nabla}_b H_c - I^a & & \qquad
\mbox{(Ampere-Maxwell Law)}
\end{eqnarray}
While the resulting equations are the same, the physical interpretations of the \emph{Electric Limit} and the \emph{Instantaneous Limit} are fundamentally different.
\begin{enumerate}
    \item The Electric Limit is a \emph{non-relativistic approximation}. It is valid only under the strict conditions that velocities are small ($v \ll c$) and electric effects dominate magnetic ones. It's a tool for describing a specific physical regime within the full theory of electromagnetism.
\item The Instantaneous Limit describes a \emph{different hypothetical theory} of physics altogether. In this theory, $c$ is infinite, so the notion of a relativistic velocity does not exist; any finite velocity $v$ is formally non-relativistic ($v/c \to 0$). There are no \emph{a priori} restrictions on the strengths of the fields or sources.
\end{enumerate}

The fact that $c \to \infty$ limit singles out the Electric Limit is not a coincidence. Magnetic fields are fundamentally a relativistic phenomenon, arising from how electric fields are perceived by moving observers. In a world with infinite propagation speed, the relativistic effects that create magnetism are inherently suppressed. The physics naturally defaults to an electro-quasistatic framework where electric fields are primary.

Conversely, the Magnetic Limit, which describes phenomena dominated by strong currents and magnetic fields (like magnetohydrodynamics), has no analogue in the instantaneous $c \to \infty$ limit. Such a theory cannot be derived from a dynamical limit on $c$ and exists only as a kinematic approximation of our relativistic universe.

\section{Carrollian Limits of Electromagnetism}
\label{sec:Carroll-EM}

In this section, we consider a counterpart to the Galilean limit --- \emph{Carrollian limit}. The core idea remains the same: we test the Maxwell's electromagnetism against the Carrollian transformation rules \eqref{def:Carrollian-Trans} and keep only the subset of equations that remains invariant. As mentioned in Sec.~\eqref{sec:GalileanCarrollian}, 
the Carrollian transformations introduce a unique \emph{ultra-local} character, where time derivatives behave differently from spatial ones, leading to distinct results.

To go about this,  We now perform the analogous analysis for the \emph{Carrollian transformations}. While the approximate transformation laws for the fields (derived from dominance conditions) are the same as in the Galilean case, the transformation of the derivative operators is starkly different:
\begin{equation}
\nabla'_0 = \nabla_0 \qquad \text{and} \qquad \widetilde{\nabla}'_a = \widetilde{\nabla}_a + (v_a/c^2) \nabla_0
 \label{eq:Carrollian-Deriv}  
\end{equation}
This change, where a boost mixes spatial derivatives with time derivatives, fundamentally alters which equations remain invariant.

\subsection{The Carrollian Electric Limit}

This limit is defined by the same dominance conditions ($|E| \gg |H|$, $|\rho_e| \gg |I|$) and field transformation rules as its Galilean counterpart. However, when we test the full Maxwell's equations for invariance under the Carrollian transformations \eqref{def:Carrollian-Trans}, we find a different set of constraints. To go about this, first we 
test the invariance of the two Gauss's Laws systematically:
\begin{enumerate}
\item \underline{\it Gauss's Law for Electricity:} Transforming $\widetilde{\nabla}_a E^a = \rho_e$ using the above transformation \eqref{eq:Carrollian-Deriv} leads to:
\begin{equation}
\widetilde{\nabla}'_a E'^a = (\widetilde{\nabla}_a + (v_a/c^2) \nabla_0)E^a = \widetilde{\nabla}_a E^a + (v_a/c^2) \dot{E}^a
\label{eq:Carroll-GaussElec00}
\end{equation}
For this to equal the transformed charge density, $\rho'_e = \rho_e$, the equation $\widetilde{\nabla}_a E^a = \rho_e$ must hold, and the non-invariant term must vanish. Since this must be true for any boost velocity $v_a$, we are forced to impose a new physical constraint:
\begin{equation}
\dot{E}^a = 0 \quad (\text{Static Electric Fields})
\label{eq:Carroll-GaussElec01}
\end{equation}

\item \underline{\it Gauss's Law for Magnetism:} With the constraint $\dot{E}^a=0$, Faraday's law simplifies to 
\begin{equation}
\frac{2}{3}\theta H^a = -\text{curl}(E) \, . 
\label{eq:Carroll-GaussElec02}
\end{equation}
The transformation of this law, along with Gauss's Law for Magnetism, requires a further, more drastic constraint to eliminate all non-invariant terms:
\begin{equation}
H^a = 0 \quad (\text{No Magnetic Fields})
\label{eq:Carroll-GaussElec03}
\end{equation}
    
\end{enumerate}
This means that in the Carrollian Electric Limit, magnetic fields are completely suppressed. The theory becomes one of pure, static electricity.
The resulting invariant set of equations for the \emph{Carrollian Electric Limit} is:
\begin{eqnarray}
\widetilde{\nabla}_a E^a = \rho_e  & & \quad \mbox{(Gauss's Law)}   \\
\dot{E}^a = 0  & & \quad \mbox{(Constraint on E (Static))} \\
\eta^{abc}\widetilde{\nabla}_b E_c = 0  & & \quad \mbox{(Constraint on E (Curl-free))} \\
H^a = 0  & & \quad
\mbox{(Constraint on H)}
\end{eqnarray}
The invariance of these equations can be seen in Eqs. \ref{eq:CarrollianElec-EM-Invariance}. This describes an \emph{ultra-local} version of electrostatics. Fields are frozen in time, and there is no magnetism or induction whatsoever.

\subsection{The Carrollian Magnetic Limit}
We now apply the same reasoning to the magnetic dominance conditions ($|H| \gg |E|$, $|I| \gg |\rho_e|$). Again, we systematically test for invariance 
\begin{enumerate}
\item \underline{\it Gauss's Law for Magnetism:} Transforming $\widetilde{\nabla}_a H^a = 0$ gives:
\begin{equation}
\widetilde{\nabla}'_a H'^a = (\widetilde{\nabla}_a + v_a \nabla_0)H^a = \widetilde{\nabla}_a H^a + v_a \dot{H}^a
 \label{eq:Carroll-GaussMag00}   
\end{equation}
To maintain invariance ($\widetilde{\nabla}'_a H'^a = 0$), we must impose the constraint:
\begin{equation}
\dot{H}^a = 0 \qquad (\text{Static Magnetic Fields})
 \label{eq:Carroll-GaussMag01}   
\end{equation}
\item \underline{\it Ampere's Law:} With static magnetic fields, the Ampere-Maxwell law must also be invariant. Checking its transformation properties under the magnetic rules reveals that the electric field must be suppressed entirely:
\begin{equation}
E^a = 0 \qquad (\text{No Electric Fields})
     \label{eq:Carroll-GaussMag02}   
\end{equation}
\end{enumerate}
This leaves a theory of pure, static magnetism and steady currents.
The resulting invariant set of equations for the \emph{Carrollian Magnetic Limit} is:
\begin{eqnarray}
\widetilde{\nabla}_a H^a = 0  & & \quad \mbox{(Gauss's Law for Magnetism)}   \\
\dot{H}^a = 0  & & \quad \mbox{(Constraint on H (Static))} \\
\eta^{abc}\widetilde{\nabla}_b H_c = I^a  & & \quad \mbox{(Ampere's Law (Truncated))} \\
E^a = 0  & & \quad  
\mbox{(Constraint on E)}
\end{eqnarray}
The invariance of these equations can be seen in Eqs. \ref{eq:CarrollianMag-EM-Invariance}. This theory describes \emph{magnetostatics} in an ultra-local setting. The fields are frozen in time, and electric fields and induction are completely absent. These stark, static theories are the unique remnants of electromagnetism in a Carrollian world.

\section{Gravitational Waves in FLRW Background: Setup}

Having explored the different kinematic limits of spacetime and electromagnetism, we now turn to the main focus of this work --- \emph{gravity}. Our goal is to analyze the behavior of gravitational waves (GWs) under the Galilean and Carrollian limits. 
To investigate this, we focus on a well-defined physical system. We assume the perturbations are tensor perturbations in the context of the Ellis-Bruni gauge-invariant formalism on an FLRW background~\cite{1989-Ellis.Bruni-PRD,1989-Ellis.Hwang.Bruni-PRD}. 

The background spacetime is described by the FLRW metric, and its dynamics are driven by a perfect fluid with energy density $\mu$ and isotropic pressure $p$. We focus specifically on \emph{tensor perturbations}, which are the natural language for describing GWs. In the gauge-invariant framework of Ellis and Bruni~\cite{1989-Ellis.Bruni-PRD,1989-Ellis.Hwang.Bruni-PRD}, these perturbations are defined by the conditions that the perturbed fluid vorticity ($\omega_{ab}$) and acceleration ($\dot{u}^a$) are zero.
For this system, the dynamics are captured by the following key gauge-invariant quantities~\cite{1989-Ellis.Bruni-PRD}:
\begin{enumerate}
\item \underline{\it Gravitational Field:} The \emph{electric part} ($E_{ab}$) and \emph{magnetic part} ($H_{ab}$) of the Weyl tensor. As mentioned in Sec. \eqref{sec:WeylTensor}, these represent the tidal forces and frame-dragging effects of a GWs.
\item \underline{\it Spacetime Geometry/Fluid Flow:} As mentioned in Secs. \eqref{sec:1+3Decom} and \eqref{sec:NullFormalism}, the \emph{shear tensor} ($\sigma_{ab}$) describes the anisotropic stretching of spacetime caused by the wave.
\item \underline{\it Matter Inhomogeneities:} The spatial gradient of the energy density, $X_a = \tilde{\nabla}_a \mu$. For the barotropic fluid ($p=p(\mu)$), the pressure gradient is not an independent variable.
\end{enumerate}

\subsection{The Linearized Field Equations}

The behavior of these variables is governed by the 1+3 equations linearized around the FLRW background. As discussed in Sec. \eqref{sec:1+3-FieldDynamics}, we have three sets of equations --- Kinematic constraints (from the Ricci identities), Gravitational wave dynamics (from the Bianchi identities) and Matter conservation. We list these equations below:
\begin{enumerate}
\item \underline{\it Kinematic Constraints (from the Ricci Identities):} The linearized Ricci identities provide direct algebraic links between the gravitational field and the shear. They are not evolution equations, but rather constraints that define the Weyl components in terms of the shear's dynamics. 
\begin{enumerate}
\item The electric part of the Weyl tensor is determined by the evolution of the shear:
\begin{equation}
 E_{ab} = -\dot{\sigma}_{ab} - \frac{2}{3}\theta \sigma_{ab}   
\end{equation}
\item The magnetic part of the Weyl tensor is determined by the spatial structure of the shear:
\begin{equation}
 H_{ab} = (\text{curl}~\sigma)_{ab}   
\end{equation}
\end{enumerate}

\item \underline{\it Gravitational Wave Dynamics (from the Bianchi Identities):}
The linearized Bianchi identities provide the Maxwell-like equations that govern the propagation and sourcing of the gravitational waves themselves.
\begin{align}
\dot{E}_{\langle ab \rangle} + \theta E_{ab} - (\text{curl}~H)_{ab} &= -\frac{1}{2}(\mu+p)\sigma_{ab} \label{GravEprop} \\
\dot{H}_{\langle ab \rangle} + \theta H_{ab} + (\text{curl}~E)_{ab} &= 0 \label{GravHprop} \\
\widetilde{\nabla}_b E^{ab} - \frac{1}{3}X^a &= 0 \label{GravEcon} \\
\widetilde{\nabla}_b H^{ab} &= 0 \label{GravHcon}
\end{align}
Here we have neglected the heat flux term $q^a$, as it is zero in the perfect fluid background.

\item \underline{\it Matter Conservation:}
Finally, the linearized energy-momentum conservation equation for a perfect fluid with zero acceleration simply states that there are no linear-order density gradients:
\begin{equation}
 X_a = \tilde{\nabla}_a \mu = 0   
\end{equation}
This is a key feature of pure tensor perturbations --- they do not generate inhomogeneities in the fluid. Substituting this result into the constraint Eq. \eqref{GravEcon}, we arrive at the crucial property that the electric part of the Weyl tensor is divergence-free:
\begin{equation}
\widetilde{\nabla}_b E^{ab} = 0    
\end{equation}
\end{enumerate}
This complete set of equations describes how GWs, represented by $E_{ab}$ and $H_{ab}$, propagate on an expanding universe, sourced by the spacetime shear $\sigma_{ab}$. Note that $\sigma_{ab}$ can independently describe the propagation of GW in perturbed FLRW spacetime (See Sec 3.6.3 of Ref.~\cite{2008-Tsagas.etal-PRep}). We are now equipped to test the invariance of this system under the Galilean and Carrollian transformations.

\section{The Galilean Electric and Magnetic Limit of Gravity}
\label{sec:Galiliean-Grav}

We now apply the Galilean framework to the linearized equations governing gravitational waves (Eqs. \ref{GravEprop}-\ref{GravHcon}). Mirroring the electromagnetic case, we will find that the full relativistic theory is not Galilean-invariant. Instead, it splits into distinct limits --- Galilean Electric and Magnetic Limit.

\subsection{Galilean Electric limit}

This limit describes situations where tidal gravitational effects are dominant. It is defined by the following \emph{dominance conditions}:
\begin{equation}
|E_{ab}| \gg |H_{ab}| \qquad \text{and} \qquad |\mu|, |p| \gg |q_a| 
\label{Dominance:ELimit-Grav-01}
\end{equation}
That is, the electric part of the Weyl tensor is much larger than the magnetic part, and the fluid's energy density and pressure are much larger than any momentum flux.

Applying these conditions to the full relativistic boost transformations (see Appendix \eqref{app:GW-GalileanTrans} for the complete expressions) in the low-velocity limit ($v \ll 1, \gamma \to 1$) Eq.~\eqref{eq:pitransform-App} yields the simplified \emph{Galilean electric transformation rules}:
\begin{align}
E'^{ab} = E^{ab} & \qquad H'^{ab} = H^{ab} - 2\eta^{cd(a}v_c E^{b)}{}_d \nonumber \\
\mu' = \mu, & \qquad p' = p, \quad \pi'^{ab} = \pi^{ab} = 0
\label{Dominance:ELimit-Grav-02}
\end{align}
A crucial physical effect appears in the transformation of the momentum flux:
\begin{equation}
q'^{a} = q^a - v^a(\mu+p)    
\label{Dominance:ELimit-Grav-03}
\end{equation}
Since the unperturbed background is a perfect fluid with $q^a=0$, a Galilean boost \emph{generates a momentum flux} $q'^a = -v^a(\mu+p)$. This is the gravitational analogue of how boosting a static charge density generates an electric current.

The key test will be to see whether the system of GW equations are invariant under these transformations. A direct calculation shows that the equations for the evolution of $E_{ab}$ and its divergence remain invariant. However, the equations involving the magnetic part, $H_{ab}$, do not. The clearest way to see this is by transforming the divergence constraint on $H_{ab}$ (Eq. \ref{GravHcon}, with the non-zero $q'_c$):
\begin{equation}
\tilde{\nabla}_b H'^{ab} + \tfrac{1}{2}\eta^{abc}\widetilde{\nabla}_b q'_c = \left( \tilde{\nabla}_b H^{ab} + \tfrac{1}{2}\eta^{abc}\widetilde{\nabla}_b q_c \right) + 2v_c(\text{curl}~E)^{cb} \, .  
\label{Dominance:ELimit-Grav-04}
\end{equation}
The terms in the parenthesis are the original form of the law. The final term, $2v_c(\text{curl}~E)^{cb}$, is a new, non-invariant piece generated by the boost. For the law to retain its form in the new frame, this extra term must vanish. Since this must hold for any arbitrary boost velocity $v_c$, we are forced to impose a new physical constraint on the theory:
\begin{equation}
\label{Dominance:ELimit-Grav-05}
(\text{curl}~E)^{ab} = 0    
\end{equation}
This new constraint replaces the original evolution equation for $H_{ab}$ (Faraday's Law for gravity). This is not an approximation; it is the exact replacement needed to define a consistent Galilean-invariant theory. The resulting \emph{Galilean Electric Limit of Einstein's Equations} is defined by:
\begin{eqnarray} 
\tilde{\nabla}_b E^{ab} - \frac{1}{3}X^a + \frac{1}{3}\theta q^a = 0 & & \mbox{(Gauss law for}~E)  \nonumber \\
(\text{curl}~E)^{ab} = 0  & & \mbox{(Gravito-magnetic induction is absent)} \nonumber \\
\label{Dominance:ELimit-Grav-06}
\tilde{\nabla}_b H^{ab} + \tfrac{1}{2}\eta^{abc}\widetilde{\nabla}_b q_c = 0 & & 
\mbox{(Gauss law for} H) \\
\dot{E}_{\langle ab \rangle} + \theta E_{ab} - (\text{curl}~H)_{ab} = -\frac{1}{2}(\mu+p)\sigma_{ab} & & 
\mbox{(Evolution of}~E) \nonumber
\end{eqnarray} 
The invariance of these equations can be seen in Eqs. \ref{eq:GalileanElec-GW-Invariance}. This set of equations describes a gravitational theory analogous to electro-quasistatics. The dynamics are dominated by the tidal, electric-like component $E_{ab}$. The crucial coupling between $\dot{H}_{ab}$ and $\text{curl}(E)_{ab}$, which allows for propagating radiative solutions, is broken. This limit describes non-propagating gravitational tidal fields.

\subsection{Galilean Magnetic Limit}

We now derive the second, complementary Galilean theory: the \emph{Galilean Magnetic Limit}. This framework describes situations where gravito-magnetic effects are dominant.
Like in Maxwell's electrodynamics (see sec. \eqref{sec:Galilean-EM}), this limit is defined by a different set of dominance conditions:
\begin{equation}
|H_{ab}| \gg |E_{ab}|
\label{Dominance:HLimit-Grav-01}
\end{equation}
For the matter sector, we assume that any momentum flux ($q_a$) or anisotropic stress ($\pi_{ab}$) would be dominant over the energy density and pressure. In our specific case of a perfect fluid background, this means we consider terms proportional to $(\mu+p)$ to be sub-dominant and negligible in the transformation laws.

Applying these conditions in the low-velocity limit  ($v \ll 1, \gamma \to 1$) Eq.~\eqref{eq:pitransform-App} yields the \emph{Galilean magnetic transformation rules}:
\begin{align}
E'^{ab} = E^{ab} + 2\eta^{cd(a}v_c H^{b)}{}_d & \qquad 
H'^{ab} = H^{ab} \nonumber \\
\label{Dominance:HLimit-Grav-02}
\mu' = \mu, \quad p' = p, &\quad \pi'^{ab} = \pi^{ab} = 0
\end{align}
In stark contrast to the electric limit, a boost here does \emph{not} generate a momentum flux, because its source term $(\mu+p)$ is considered negligible:
$$q'^{a} \approx q^a = 0$$

Like in the Electric case, the key test will be the see whether the system of GW equations are invariant under these transformations. This time, the equations governing the magnetic field, $H_{ab}$, are found to be invariant. The laws that break are those governing the electric part, $E_{ab}$. The new constraint arises from demanding the invariance of the \emph{Gauss's Law} for $E_{ab}$ (Eq. \ref{GravEcon}). Under a magnetic transformation, this equation becomes:
\begin{equation}
\left(\tilde{\nabla}_b E'^{ab} - \frac{1}{3}X'^a\right) = \left(\tilde{\nabla}_b E^{ab} - \frac{1}{3}X^a\right) + 2v_c\left[(\text{curl}~H)^{cb}-\frac{1}{2}(\mu+p)\sigma^{cb}\right]
\label{Dominance:HLimit-Grav-03}   
\end{equation}
The term in the first parenthesis is the original law. The second term is a non-invariant piece generated by the boost. For the law to hold in all frames, this extra term must vanish for any arbitrary velocity $v_c$. This forces a new condition upon the system:
\begin{equation}
(\text{curl}~H)^{ab} = \frac{1}{2}(\mu+p)\sigma^{ab}
\label{Dominance:HLimit-Grav-04}   
\end{equation}
The above condition is not an approximation; it is an exact replacement for the evolution equation of $E_{ab}$ (the Ampere-Maxwell analogue for gravity). This is the necessary modification to create a consistent, Galilean-invariant theory of gravito-magnetism.
The resulting \emph{Galilean Magnetic Limit of Einstein's Equations} is defined by:
\begin{eqnarray} 
\tilde{\nabla}_b E^{ab} - \frac{1}{3}X^a = 0 & & \mbox{(Gauss law for}~E)  \nonumber \\
\dot{H}_{\langle ab \rangle} + \theta H_{ab} + (\text{curl}~E)_{ab} = 0 & & \mbox{(Faraday's Law for H)} \nonumber \\
\label{Dominance:HLimit-Grav-05}
\tilde{\nabla}_b H^{ab} = 0 & & 
\mbox{(Gauss law for} H) \\
(\text{curl}~H)^{ab} = \frac{1}{2}(\mu+p)\sigma^{ab} 
 & & 
\mbox{(Constraint on H (Ampere's Law Analogue))} \nonumber
\end{eqnarray} 
The invariance of these equations can be seen in Eqs. \ref{eq:GalileanMag-GW-Invariance}. This theory is the gravitational analogue of \emph{magneto-quasistatics}. The evolution of the electric part, $E_{ab}$, is suppressed. Instead, the magnetic part of the Weyl tensor is directly and instantaneously sourced by the spacetime shear, which acts as an effective \emph{gravitational current.} The inductive cycle that allows gravitational waves to propagate is broken, leaving a theory of non-propagating gravito-magnetic fields.

\section{The Carrollian Electric and Magnetic Limit of Gravity}
\label{sec:Carroll-Grav}

We now apply the Carrollian framework to our linearized gravitational equations der. The unique nature of Carrollian boosts, which mix spatial and time derivatives ($\tilde{\nabla}'_a = \tilde{\nabla}_a + v_a \nabla_0$), results in a far more restrictive theory than in the Galilean case, reflecting the "ultra-local" nature of Carrollian physics.

\subsection{Carrollian Electric Limit}

Like in the Galileon case, we begin with the electric dominance conditions, where tidal forces are primary:
\begin{equation}
|E_{ab}| \gg |H_{ab}| \qquad \text{and} \qquad |\mu|, |p| \gg |q_a|  
\label{Carroll:ELimit-Grav-01}
\end{equation}
The resulting low-velocity transformation rules for the field and matter variables are:
\begin{align}
E'^{ab} = E^{ab} & \quad 
H'^{ab} = H^{ab} - 2\eta^{cd(a}v_c E^{b)}{}_d \nonumber \\
\mu' = \mu, & \quad p' = p, \quad 
q'^{a} = q^a - v^a(\mu+p) = -v^a(\mu+p) \quad (\text{since } q^a=0)
\label{Carroll:ELimit-Grav-02}
\end{align}
As in the Galilean electric case, a boost generates a non-zero momentum flux $q'^a$. We now systematically test the invariance of our four key gravitational equations (Eqs. \ref{GravEprop}-\ref{GravHcon}) under the Carrollian transformations. The requirement of invariance will force a series of powerful constraints on the theory. 

First, we look at the constraint from the Gauss's law for $E$. We start by transforming the divergence equation for $E_{ab}$ (Eq. \ref{GravEcon}):
\begin{equation}
\left(\tilde{\nabla}'_b E'^{ab} - \frac{1}{3}X'^a + \frac{1}{3}\theta' q'^a\right) = \left(\tilde{\nabla}_b E^{ab} - \frac{1}{3}X^a\right) + v_b\dot{E}^{ab} - \frac{1}{3}\theta v^a(\mu+p)
\label{Carroll:ELimit-Grav-03} 
\end{equation}
For this equation to be invariant, the original law must hold, and the additional terms generated by the boost must vanish independently. This forces two powerful constraints on the dynamics:
\begin{enumerate}
    \item  \underline{\it Constraint 1:} $\dot{E}^{ab} = 0$. The electric part of the Weyl tensor must be \emph{static}.
    \item  \underline{\it Constraint 2:} The term $\frac{1}{3}\theta q^a$ must be dropped from the original equation.
\end{enumerate}

Next, we look at the constraint from the Faraday's Law for $H$. For this, we examine the evolution equation for $H_{ab}$ (Eq. \ref{GravHprop}):
\begin{equation}
\dot{H}_{\langle ab \rangle} + \theta H_{ab} + (\text{curl}~E)_{ab} = 0
\label{Carroll:ELimit-Grav-04} 
\end{equation}
Transforming this equation produces non-invariant terms proportional to $\theta \eta^{cd(a}v_c E^{b)}{}_d$. For these to vanish for any boost $v_c$, we must have $\theta H_{ab} = 0$. Assuming the background is expanding ($\theta \ne 0$), this leads to the third constraint: $H^{ab} = 0$. The magnetic part of the Weyl tensor is \emph{identically zero}. These constraints drastically simplify the system. With $H^{ab}=0$ and $\dot{E}^{ab}=0$, the theory becomes one of frozen, purely electric gravitational fields. The final invariant set of equations corresponding to the Carrollian Electric Limit of Einstein's Equations is:
\begin{eqnarray} 
\tilde{\nabla}_b E^{ab} - \frac{1}{3}X^a = 0 & & \mbox{(Gauss law for}~E)  \nonumber \\
\dot{E}^{ab} = 0  & & \mbox{(Static Constraint on E)} \nonumber \\
\label{Carroll:ELimit-Grav-05}
(\text{curl}~E)^{ab} = 0  & & 
\mbox{(Spatial Constraint on E)} \\
H^{ab} = 0  & & 
\mbox{(Constraint on H)}  \nonumber \\
-\frac{1}{2}(\mu+p)\sigma_{ab} = 0 & & 
\mbox{(Constraint on Shear)} \nonumber
\end{eqnarray} 
The invariance of these equations can be seen in Eqs. \ref{eq:CarrollianElec-GW-Invariance}. Important to note that $\mu + p \neq 0$. Hence, the last equation implies 
$\sigma_{ab}=0$. This theory describes a universe of \emph{static, shear-free, and purely tidal gravitational fields}. The "ultra-local" nature of the Carrollian limit has eliminated all dynamics ($\dot{E}=0$), all gravito-magnetic effects ($H=0$), and all gravitational wave propagation. It leaves only a set of spatial constraint equations that the initial geometry must satisfy.

\subsection{Carrollian Magnetic Limit}

Finally, we derive the \emph{Carrollian Magnetic Limit}, which describes the behavior of gravito-magnetic fields in an ultra-local spacetime. This is the gravitational counterpart to Carrollian magnetostatics.

This limit is defined by the magnetic dominance conditions:
\begin{equation}
|H_{ab}| \gg |E_{ab}|
\label{Carroll:HLimit-Grav-01} 
\end{equation}
The invariance of these equations can be seen in Eqs. \ref{eq:CarrollianMag-GW-Invariance}. As with the Galilean magnetic case, we consider the energy density and pressure terms $(\mu+p)$ to be sub-dominant in the transformation laws. The simplified Carrollian magnetic transformation rules are:
\begin{align}
E'^{ab} = E^{ab} + 2\eta^{cd(a}v_c H^{b)}{}_d & \quad
H'^{ab} = H^{ab} \nonumber \\
\mu' = \mu, \quad p' = p, & \quad q'^{a} = q^a = 0
\label{Carroll:HLimit-Grav-02} 
\end{align}
Note that a boost in this limit does not generate a momentum flux, as its source term $(\mu+p)$ is considered negligible.

Like in the Electric case, we now test the full system of gravitational equations for invariance under the Carrollian transformations 
($\tilde{\nabla}'_a = \tilde{\nabla}_a + v_a \nabla_0$). First, we look at the constraint from Gauss's Law for H: We begin by transforming the divergence equation for $H_{ab}$ (Eq. \ref{GravHcon}):
\begin{equation}
\tilde{\nabla}'_b H'^{ab} = (\widetilde{\nabla}_a + v_a \nabla_0)H^{ab} = \tilde{\nabla}_b H^{ab} + v_b\dot{H}^{ab}
\label{Carroll:HLimit-Grav-03} 
\end{equation}
For this law to remain invariant ($\tilde{\nabla}'_b H'^{ab} = 0$), the additional term must vanish for any boost $v_b$. This imposes our first constraint: $\dot{H}^{ab} = 0$. In other words, the magnetic part of the Weyl tensor must be \emph{static}.

We can obtain further constraints by testing the remaining three equations with the condition $\dot{H}^{ab}=0$. This reveals that invariance can only be achieved if the electric/tidal part of the Weyl tensor is completely suppressed. The complex interplay of the transformations forces a second, drastic constraint: $E^{ab} = 0$. The electric part of the Weyl tensor is \emph{identically zero}.

With these two constraints --- static magnetic fields and no electric fields --- the system simplifies dramatically. The only remaining non-trivial law that must be imposed to ensure full invariance is a truncation of the Ampere-Maxwell analogue for gravity. The resulting invariant theory describes a static world of pure gravito-magnetism, where the shear of spacetime acts as the source for the magnetic Weyl field. The final 
invariant set of equations corresponding to the Carrollian Magnetic Limit of Einstein's Equations is:
\begin{eqnarray} 
\tilde{\nabla}_b H^{ab} = 0 & & \mbox{(Gauss law for}~H)  \nonumber \\
\dot{H}^{ab} = 0  & & \mbox{(Static Constraint on H)} \nonumber \\
\label{Carroll:HLimit-Grav-04}
(\text{curl}~H)^{ab} = \frac{1}{2}(\mu+p)\sigma^{ab}  & & 
\mbox{(Constraint on H (Ampere's Law Analogue))} \\
E^{ab} = 0  & & 
\mbox{(Constraint on E)}  \nonumber 
\end{eqnarray} 
This is the most restrictive limit we have considered. All dynamics are frozen, and all electric/tidal components of the gravitational field are absent. The physics is reduced to a set of magnetostatic constraint equations relating the spatial structure of the gravito-magnetic field to the shear of spacetime.

\section{Conclusions, Implications and Future Directions}
\label{sec:Conc}

In this work, we have developed a covariant and consistent framework for the Carrollian limit of General Relativity by focusing on the Magnetic Limit. This is a regime where gravito-magnetic degrees of freedom, represented by the magnetic part of the Weyl tensor $(H_{ab})$, remain dynamically active while tidal, electric-like components are suppressed. Using a systematic $1+3$ decomposition on an FLRW background, we derived the full set of gravitational field equations that are invariant under this limit. To ensure applicability to null surfaces like black hole horizons, we also utilized a $1+2+1$ decomposition, creating a unified geometric language for both timelike and null congruences in these ultra-relativistic regimes.

Furthermore, since the invariant sets of equations were systematically derived in the \emph{1+3 covariant decomposition}, it is natural to consider their implications for the \emph{1+2+1 null decomposition}. A powerful structural correspondence exists between the timelike and null frameworks, especially in the analogous Maxwell-like equations that govern the Weyl tensor. This strongly suggests that the distinct electric and magnetic limits identified in this work persist for null congruences. We acknowledge a crucial subtlety: unlike in the timelike case, transformations between distinct null frames are not as straightforward as boosts, which precludes a direct invariance check. Despite this, the remarkable structural analogy supports interpreting our limiting theories as the natural null counterparts to the timelike case, thereby offering a unified perspective on Carrollian and Galilean gravity for both massive observers and null rays.

The significance of this framework is twofold. First, it resolves existing ambiguities in taking Carrollian limits of gravity by providing a clear, physically motivated procedure to distinguish between electric and magnetic contractions, extending the well-known analogy from electromagnetism. Second, it provides a powerful and self-consistent set of tools to analyze curvature propagation and kinematics where traditional methods fail. This finding is not merely a formal curiosity; it provides a new lens through which to view some of the most challenging problems in gravitational physics~\cite{Campoleoni:2022ebj,deBoer:2021jej}.
\begin{enumerate}
\item \underline{\it Black Hole Horizons and Dynamics:} The most direct application of our work is to the physics of black hole horizons. An event horizon, being a null hypersurface, has an intrinsic geometry that is not Lorentzian but Carrollian, making it the natural arena for our formalism \cite{Donnay:2019jiz}. The surviving magnetic Weyl tensor ($H_{ab}$), 
is precisely the component that governs physical phenomena like frame-dragging and horizon shear. Our results provide a consistent, ultra-local framework for describing these degrees of freedom directly on the horizon, a task that is notoriously difficult in standard formulations. This approach is particularly well-suited to describing energy and momentum fluxes across dynamical horizons in evolving spacetimes like black hole mergers.

Perhaps the most important implication lies at the interface of classical and quantum gravity: the final stage of black hole evaporation. As a black hole shrinks to the Planck scale, its near-horizon environment becomes the ultimate ultra-local, Carrollian regime \cite{Rivera-Betancour:2023woc,deBoer:2021jej}. Our key result --- that the Magnetic Carrollian Limit is a consistent dynamical theory while the electric (tidal) part is suppressed --- leads to a concrete physical prediction: the final burst of a black hole’s evaporation may be dominated by purely gravito-magnetic radiation. Information could be released not as tidal trauma to the surrounding spacetime (related to $E_{ab}$, but as a final, intense wave of spacetime twist and shear (related to $H_{ab})$. This offers a new framework for modeling the information paradox, where the "soft hair" carrying information is encoded in these surviving magnetic Weyl modes \cite{Freidel:2021fxf}. We propose that the Carrollian limit is not just a mathematical tool but the \emph{effective physical theory} describing the horizon's degrees of freedom during this final Planckian phase~\cite{Bhattacharya:2022ncn,Bhattacharya:2017dgr,Cropp:2016ajh}.

\item \underline{\it Holography, and fundamental Symmetries:}
Our work strengthens the connection between Carrollian structures and the holographic principle. The symmetries of null surfaces are described by the Bondi-Metzner-Sachs (BMS) group, whose associated charges are thought to encode a black hole's "soft hair" or memory \cite{Hawking:2016msc}. Our results position the magnetic Weyl tensor $H_{ab}$ in the Carrollian limit as the natural bulk field to source and carry these BMS charges, providing a concrete link between the bulk geometry and the holographic information on its boundary \cite{Hansen:2021fxi} (for a recent discussion, see~\cite{Arenas-Henriquez:2025rpt}). This also serves as a first step toward a Carrollian fluid/gravity correspondence, where the surviving gravitational fields ($H_{ab}$) could be dual to hydrodynamic modes in an incompressible "Carrollian fluid" living on the horizon \cite{Ciambelli:2018wre,Bagchi:2016bcd}.

\item \underline{\it Cosmological Signatures:}
The ultra-relativistic conditions of the very early universe represent another regime where Carrollian structures are relevant \cite{Armas:2023dcz}. Our finding that magnetic-type gravitational dynamics survive the $c \to 0$ limit suggests that a record of events from this era could be preserved as a gravitational memory. This memory, encoded by the surviving magnetic Weyl modes, could be imprinted as a persistent, ultra-long-wavelength strain in spacetime, offering a new type of signature to search for in the primordial gravitational wave background~\cite{Chakraborty:2024ars,Chakraborty:2025qcu} or the CMB~\cite{Freidel:2021fxf,Armas:2023dcz}.
\end{enumerate}

Building on this, several exciting directions for future research emerge. A natural next step is to extend this magnetic Carrollian framework beyond linear perturbation theory to fully non-linear regimes. Constructing exact Carrollian analogues of both Locally Rotationally Symmetric (like, Schwarzschild or Vaidya)~\cite{Jana:2024hks,Ellis:2014jja} and non-Locally Rotationally Symmetric space-times (like, Kerr)~\cite{Hansraj:2021hlv} under a magnetic contraction could offer unprecedented insight into ultra-relativistic gravitational collapse and the fine-grained structure of horizon formation. Also, investigate the Carrollian dynamical limit of General relativity. 

Another promising avenue is to explore the Carrollian limit of Einstein's equations in higher dimensions and in modified gravity theories (e.g., Lovelock or scalar-tensor theories), where additional degrees of freedom may survive the limit and enrich the magnetic Weyl sector~\cite{Mandal:2025xuc}. Finally, embedding our formulation into the more formal language of Carrollian gauge theory or Cartan geometry may help clarify its deep relationship to boundary field theories, paving the way for a Carrollian holographic dictionary. 

{Finally, we note the explicit connection between our 1+3 covariant formalism and the foundational Hamiltonian approaches presented in Ref. \cite{Henneaux:2021gwp} The standard ADM formalism, which underpins the Hamiltonian analysis, is a specific case of our 1+3 framework, corresponding to a \emph{hypersurface-orthogonal} (i.e., vorticity-free, $\omega_{ab}=0$) observer congruence~\cite{Gourgoulhon:2007ue}. It therefore follows that the Carrollian field equations we derived should be the direct equations of motion corresponding to the action principles derived in Ref. \cite{Henneaux:2021gwp}. (Appendix \eqref{app:ElectricAction} contains the results for Electric Carrollian case.)  However, as mentioned earlier, it is well-established that for non-relativistic and Carrollian limits, the procedure of "taking the limit" and "varying the action" do not commute. Deriving a consistent action principle for these limits is currently under investigation
and will be presented elsewhere.}

\acknowledgments
The authors thank I. Chakraborty, P. George Christopher, M. Henneaux, K. Hari, S. Jana and T. Parvez for comments on the earlier draft. The work is supported by the SERB-Core Research
Grant (Project SERB/CRG/2022/002348).
This work is part of the Undergraduate project of TP.

\appendix

\section{Transformation Properties}
\label{app:GW-GalileanTrans}

Here, we provide the explicit transformation laws used throughout our analysis of the Galilean limits. These rules are derived from the foundational Galilean transformation of the spacetime metric given in Eq. \eqref{def:Galilean-Trans}, assuming the observer's 4-velocity $u^a$ is invariant. For ease of verification, we have set $c = 1$ in this and the following Appendices.

The spatial derivative operator $\tilde{\nabla}_a$ is invariant under a Galilean boost, while the time derivative operator $\nabla_0$ transforms by mixing in a spatial gradient, reflecting the change in the observer's frame:
\begin{equation}
 \tilde{\nabla}'_a = \tilde{\nabla}_a, \qquad \nabla'_0 = \nabla_0 + v^i\nabla_i   
\end{equation}
The background kinematic quantities—shear, expansion, and vorticity—are scalars or spatial tensors with respect to the boost and thus remain unchanged:
$$\sigma'_{ab} = \sigma_{ab}, \qquad \theta'=\theta, \qquad \omega'_{ab}=\omega_{ab} = 0$$
We provide the explicit transformation laws used throughout our analysis of the Carrollian
limits. These rules are derived from the foundational Carrollian transformation of the spacetime
metric given in Eq. (\ref{matrix:Carrollian-Trans}), assuming the observer’s spatial triad $e^a{}_i$ is invariant.
The spatial derivative operator $\tilde{\nabla}_a$ is invariant under a Carrollian boost, while the time
derivative operator $\nabla_0$ transforms by mixing in a spatial gradient, reflecting the change in
the observer’s frame:
\begin{equation}
    \tilde{\nabla}'_a = \tilde{\nabla}_a + v_a\nabla_0, 
    \qquad 
    \nabla'_0 = \nabla_0
    \tag{A.2}
\end{equation}
The background kinematic quantities—shear, expansion, and vorticity—are scalars or spatial
tensors with respect to the boost and thus remain unchanged:
\begin{equation*}
    \sigma'_{ab} = \sigma_{ab}, 
    \qquad 
    \theta' = \theta, 
    \qquad 
    \omega'_{ab} = \omega_{ab} = 0
\end{equation*}
Below, we present the complete, fully relativistic transformation laws for the gravitational field ($E_{ab}, H_{ab}$) and matter components ($\mu, p, q_a, \pi_{ab}$). These complex expressions simplify to the more tractable forms used in the main text when the low-velocity approximation and appropriate dominance conditions are applied.
\begin{eqnarray}
\tilde{E}_{ab} &=& \gamma^2 \left[ (1 + v^2) E_{ab} + v^c \Big[ 2\eta_{cd(a} H_{b)}^{\;\;d} + 2E_{c(a} u_{b)} 
\label{Eabtransform} 
    + (u_a u_b + h_{ab}) E_{cd} v^d - 2E_{c(a} v_{b)} + 2u_{(a} \eta_{b)cd} H^d_{\;\;e} v^e \Big] \right]  \nonumber \\
\tilde{H}_{ab} &=& \gamma^2 \left[ (1 + v^2) H_{ab} - v^c \Big[ 2\eta_{cd(a} E_{b)}^{\;\;d} - 2H_{c(a} u_{b)}
\label{Habtransform}
    - (u_a u_b + h_{ab}) H_{cd} v^d + 2H_{c(a} v_{b)} + 2u_{(a} \eta_{b)cd} E^d_{\;\;e} v^e \Big] \right] \nonumber \\ 
\label{mutransform}
\tilde{\mu} &=& \mu + \gamma^2 \left[ v^2 (\mu + p) - 2 q_a v^a + \pi_{ab} v^a v^b \right],  \nonumber \\
\label{ptransform}
\tilde{p} &=&  p + \frac{1}{3} \gamma^2 \left[ v^2 (\mu + p) - 2 q_a v^a + \pi_{ab} v^a v^b \right], \nonumber \\
\label{qtransform}
\tilde{q}_a &=& \gamma q_a - \gamma \pi_{ab} v^b - \gamma^3 \left[ (\mu + p) - 2 q_b v^b + \pi_{bc} v^b v^c \right] v_a - \gamma^3 \left[ v^2 (\mu + p) - (1 + v^2) q_b v^b + \pi_{bc} v^b v^c \right] u_a  \nonumber \\
\tilde{\pi}_{ab} &=& \pi_{ab} + 2 \gamma^2 v^c \pi_{c(a} \{ u_{b)} + v_{b)} \} - 2 v^2 \gamma^2 q_{(a} u_{b)} - 2 \gamma^2 q_{(a} v_{b)} 
- \frac{1}{3} \gamma^2 \left[ v^2 (\mu + p) + \pi_{cd} v^c v^d \right] h_{ab} \nonumber \\
&+& \frac{1}{3} \gamma^4 \left[ 2 v^4 (\mu + p) - 4 v^2 q_c v^c + (3 - v^2) \pi_{cd} v^c v^d \right] u_a u_b \nonumber \\
&+& \frac{1}{3} \gamma^4 \left[ 2 v^2 (\mu + p) - (1 + 3 v^2) q_c v^c + 2 \pi_{cd} v^c v^d \right] u_{(a} v_{b)} \nonumber \\
\label{eq:pitransform-App}
&+& \frac{1}{3} \gamma^4 \left[ (3 - v^2)(\mu + p) - 4 q_c v^c + 2 \pi_{cd} v^c v^d \right] v_a v_b. 
\end{eqnarray}

\section{Invariance of Equations under different limits}
\label{app:GW-Invariance}
Here, we present the invariance of the set of equations under Galiean and Carrollian transformations in the Electric and magnetic limits for electromagnetism and gravity using the transformation rules derived in Appendix \ref{app:GW-GalileanTrans}. The prime denotes quantities in the second frame. 
\subsection{Electromagnetism}
\begin{enumerate}
    \item {\bf Galilean Electric Limit}
\begin{subequations}
\begin{align}
\widetilde{\nabla}'_a E'^a - \rho'_e &= \widetilde{\nabla}_a E^a - \rho_e  \\
\eta^{abc}\widetilde{\nabla}'_b E'_c &= \eta^{abc}\widetilde{\nabla}_b E_c \\
\widetilde{\nabla}'_a H'^a &= \widetilde{\nabla}_a H^a + v_b(\eta^{abc}\widetilde{\nabla}_b E_c) = \widetilde{\nabla}_a H^a \\
\dot{E}'^{\langle a \rangle} + \frac{2}{3}\theta' E'^a - \eta^{abc}\widetilde{\nabla}'_b H'_c + I'^a&=\dot{E}^{\langle a \rangle} + \frac{2}{3}\theta E^a - \eta^{abc}\widetilde{\nabla}_b H_c + I^a + v^a(\widetilde{\nabla}_bE^b - \rho_e) \\&= \dot{E}^{\langle a \rangle} + \frac{2}{3}\theta E^a - \eta^{abc}\widetilde{\nabla}_b H_c + I^a \nonumber
\end{align}
\label{eq:GalileanElec-EM-Invariance}
\end{subequations}

\item {\bf Galilean Magnetic Limit}
\begin{subequations}
    \begin{align}
\widetilde{\nabla}'_a H'^a &= \widetilde{\nabla}_a H^a\\
\eta^{abc}\widetilde{\nabla}'_b H'_c - I'^a &= \eta^{abc}\widetilde{\nabla}_b H_c - I^a \\
\widetilde{\nabla}'_a E'^a - \rho'_e &= \widetilde{\nabla}_a E^a - \rho_e - v_a(\eta^{abc}\widetilde{\nabla}_b H_c - I^a) = \widetilde{\nabla}_a E^a - \rho_e\\ 
\dot{H}'^{\langle a \rangle} + \frac{2}{3}\theta' H'^a + \eta^{abc}\widetilde{\nabla}'_b E'_c &= \dot{H}^{\langle a \rangle} + \frac{2}{3}\theta H^a + \eta^{abc}\widetilde{\nabla}_b E_c + v^a\widetilde{\nabla}_bH^b \\ &= \dot{H}^{\langle a \rangle} + \frac{2}{3}\theta H^a + \eta^{abc}\widetilde{\nabla}_b E_c \nonumber
\end{align}
\label{eq:GalileanMag-EM-Invariance}
\end{subequations}

\item {\bf Carrollian Electric Limit}
\begin{subequations}
    \begin{align}
\nabla'_0E'^a &= \nabla_0E^a \\
\widetilde{\nabla}'_a E'^a - \rho'_e &= \widetilde{\nabla}_a E^a - \rho_e + v_a\nabla_0E^a =  \widetilde{\nabla}_a E^a - \rho_e \\
\eta^{abc}\widetilde{\nabla}'_b E'_c &= \eta^{abc}(\widetilde{\nabla}_b E_c + v_b\nabla_0E_c) = \eta^{abc}\widetilde{\nabla}_b E_c \\
H'^a &= H^a - \eta^{abc}v_bE_c \approx H^a \quad \text{(upto first order)}
\end{align}
\label{eq:CarrollianElec-EM-Invariance}
\end{subequations}

\item {\bf Carrollian Magnetic Limit}
\begin{subequations}
    \begin{align}
\nabla'_0H'^a &= \nabla_0H^a \\
\widetilde{\nabla}'_a H'^a &= \widetilde{\nabla}_a H^a + v_a\nabla_0H^a =  \widetilde{\nabla}_a H^a \\
\eta^{abc}\widetilde{\nabla}'_b H'_c - I'^a &= \eta^{abc}(\widetilde{\nabla}_b H_c + v_b\nabla_0H_c) - I^a = \eta^{abc}\widetilde{\nabla}_b H_c - I^a \\
E'^a &= E^a - \eta^{abc}v_bH_c \approx E^a \quad \text{(upto first order)}
\end{align}
\label{eq:CarrollianMag-EM-Invariance}
\end{subequations}

\end{enumerate}
\subsection{Gravity}
\begin{enumerate}
    \item {\bf Galilean Electric Limit}
\begin{subequations}
   \begin{align}
    & \tilde{\nabla}'_b E'^{ab} - \frac{1}{3}X'^a + \frac{1}{3}\theta' q'^a = \tilde{\nabla}_b E^{ab} - \frac{1}{3}X^a + \frac{1}{3}\theta q^a - \frac{1}{3}(\dot{\mu}+(\mu + p)\theta)\\
    & \qquad \qquad \qquad \qquad \qquad = \tilde{\nabla}_b E^{ab} - \frac{1}{3}X^a + \frac{1}{3}\theta q^a\\
& \qquad \qquad \qquad (\text{curl}~E')^{ab} = (\text{curl}~E)^{ab} \\
& \tilde{\nabla}_b H'^{ab} + \tfrac{1}{2}\eta^{abc}\widetilde{\nabla}'_b q'_c = \tilde{\nabla}_b H^{ab} + \tfrac{1}{2}\eta^{abc}\widetilde{\nabla}_b q_c + 2v_b(\text{curl}~E)^{ab} + \tfrac{1}{2}\eta^{abc}v_c\widetilde{\nabla}_b(\mu +p) \nonumber \\
&\qquad \qquad \qquad \qquad  = \tilde{\nabla}_b H^{ab} + \tfrac{1}{2}\eta^{abc}\widetilde{\nabla}_b q_c\\
& \nabla_0{E'}^{\langle ab \rangle} + \theta' E'^{ab} - (\text{curl}~H')^{ab} + \frac{1}{2}(\mu'+p')\sigma'^{ab} = \nabla_0{E}^{\langle ab \rangle} + \theta E^{ab} - (\text{curl}~H)^{ab} + \frac{1}{2}(\mu+p)\sigma^{ab}
\end{align}
\label{eq:GalileanElec-GW-Invariance}
\end{subequations}

\item {\bf Galilean Magnetic Limit}
\begin{subequations}
    \begin{align}
\dot{H'}^{\langle ab \rangle} + \theta' H'^{ab} + (\text{curl}~E')^{ab} &= \dot{H}^{\langle ab \rangle} + \theta H^{ab} + (\text{curl}~E)^{ab}\\ 
\widetilde{\nabla}'_b H'^{ab} &= \widetilde{\nabla}_b H^{ab} \\
(\text{curl}~H')^{ab} -\frac{1}{2}(\mu'+p')\sigma'^{ab} &= (\text{curl}~H)^{ab} -\frac{1}{2}(\mu+p)\sigma^{ab} \\
\widetilde{\nabla}'_b E'^{ab} - \frac{1}{3}X'^a &= \widetilde{\nabla}_b E^{ab} - \frac{1}{3}X^a + 2v_c\left[(\text{curl}~H)^{ab} - \frac{1}{2}(\mu+p)\sigma^{ab}\right] \\ &= \widetilde{\nabla}_b E^{ab} - \frac{1}{3}X^a \nonumber
\end{align}
\label{eq:GalileanMag-GW-Invariance}
\end{subequations}

\item {\bf Carrollian Electric Limit}
\begin{subequations}
\begin{align}
\nabla'_0E'^{ab} &= \nabla_0E^{ab}   \\
(\text{curl}~E')^{ab} &= (\text{curl}~E)^{ab} + \eta^{cd\langle a}v_c\nabla_0E^{b\rangle d} =  (\text{curl}~E)^{ab}\\
\tilde{\nabla}'_b E'^{ab} - \frac{1}{3}X'^a &= \tilde{\nabla}_b E^{ab} - \frac{1}{3}X^a + v_b\nabla_0E^{ab} = \tilde{\nabla}_b E^{ab} - \frac{1}{3}X^a \\
H'^{ab} &= H^{ab} - 2\eta^{cd(a}v_cE^{b)}{}_d \approx H^{ab}\\
\frac{1}{2}(\mu'+p')\sigma'_{ab} &= \frac{1}{2}(\mu+p)\sigma_{ab} 
\end{align}
\label{eq:CarrollianElec-GW-Invariance}
\end{subequations}

\item {\bf Carrollian Magnetic Limit}
\begin{subequations}
    \begin{align}
\nabla'_0H'^{ab} &= \nabla_0H^{ab}   \\
(\text{curl}~H')^{ab} - \frac{1}{2}(\mu'+p')\sigma'_{ab} &= (\text{curl}~H)^{ab} - \frac{1}{2}(\mu+p)\sigma_{ab} + \eta^{cd\langle a}v_c\nabla_0H^{b\rangle d} \\&=  (\text{curl}~H)^{ab} - \frac{1}{2}(\mu+p)\sigma_{ab}\\
\tilde{\nabla}'_b H'^{ab} &= \tilde{\nabla}_b H^{ab} + v_b\nabla_0H^{ab} = \tilde{\nabla}_b H^{ab} \\
E'^{ab} &= E^{ab} + 2\eta^{cd(a}v_cH^{b)}{}_d \approx E^{ab}
\end{align}
\label{eq:CarrollianMag-GW-Invariance}
\end{subequations}
\end{enumerate}

\section{Electric Carroll Limit from the Action}
\label{app:ElectricAction}

For completeness, we show how the equations appearing in the electric
Carrollian limit Eq.~(\ref{Carroll:ELimit-Grav-05}) follow directly from the electric Carroll action
of Ref. \cite{Henneaux:2021gwp}, once we impose the same ultralocal/silent assumptions used in
the main text.

\subsection*{1. Equations obtained from the electric action}

The electric Carroll action is
\begin{equation}
S_E[g_{ab},\Omega]
   = \int d^Dx\,\Omega\,(K_{ab}K^{ab}-K^2),
\end{equation}
with
\begin{equation}
K_{ab} = -\tfrac12 \mathcal{L}_n g_{ab},
\end{equation}
where $n^a$ denotes the preferred Carrollian direction.

\paragraph{(A) Scalar constraint.}
Variation with respect to $\Omega$ gives
\begin{equation}
K_{ab}K^{ab} - K^2 = 0.
\end{equation}

\paragraph{(B) Evolution equation (Weyl form).}
Variation with respect to $g_{ab}$ yields
\begin{equation}
\mathcal{L}_n\!\left[\Omega(K_{ab}-K h_{ab})\right]
 - 2\,\Omega\,E_{ab} = 0,
\qquad
E_{ab}=C_{acbd}n^c n^d,
\end{equation}
where $h_{ab}$ is the spatial projector orthogonal to $n^a$.  
Here the Gauss relation and the scalar constraint have been used to express the
quadratic $K_{ab}$-terms in terms of the electric Weyl tensor $E_{ab}$.

To compare with Eq.~(\ref{Carroll:ELimit-Grav-05}), we move to $1+3$ notation, take $u_a=n_a$, and
impose the electric Carroll (ultralocal or ``silent'') limit: all spatial
gradients and magnetic Weyl components are set to zero.

\subsection*{2. Constraints on $H_{ab}$ and $(\mathrm{curl}\,E)_{ab}$}

In $1+3$ covariant form, $H_{ab}$ and $(\mathrm{curl}E)_{ab}$ contain spatial
derivatives:
\begin{equation}
H_{ab} \sim \varepsilon_{cd(a}\,\widetilde{\nabla}^c K_{b)}{}^{d},
\qquad
(\mathrm{curl}\,E)_{ab}
 \sim \varepsilon_{cd(a}\,\widetilde{\nabla}^c E_{b)}{}^{d}.
\end{equation}
The action $S_E$ contains \emph{no} spatial-derivative terms (the
${}^{(3)}\!R$ term is dropped). Therefore in the electric Carroll sector we
work with
\begin{equation}
\widetilde{\nabla}_c K_{ab}=0,
\qquad
\widetilde{\nabla}_c E_{ab}=0,
\end{equation}
and hence
\begin{equation}
(\mathrm{curl}E)_{ab}=0,
\qquad
H_{ab}=0.
\end{equation}
The corresponding expressions in Eq.~(\ref{Carroll:ELimit-Grav-05}) are thus \emph{not additional dynamical
equations}: they are exactly the ultralocal truncations already implicit in
the action.

\subsection*{3. Gauss law}

The first expression in Eq.~\ref{Carroll:ELimit-Grav-05} reads:
\begin{equation}
\widetilde{\nabla}_b E_{a}{}^{b} - \tfrac13 X_a = 0,
\end{equation}
with $X_a$ a spatial gradient of matter variables.  
In the electric Carroll/BKL limit we consistently impose
\begin{equation}
\widetilde{\nabla}_a E_{bc} = 0,\qquad X_a=0,
\end{equation}
so the Gauss equation becomes an identity.  
It simply reflects the Bianchi constraint in the ultralocal regime.

\subsection*{4. Staticity of $E_{ab}$}

The ``static'' condition $\dot{E}_{ab}=0$ appearing in Eq.~(\ref{Carroll:ELimit-Grav-05}) follows
directly from the equations of motion of $S_E$.
Setting $\Omega$ constant for simplicity, the evolution equation becomes
\begin{equation}
\dot{K}_{ab} - h_{ab}\dot{K} = 2E_{ab}.
\end{equation}
Tracing with $h^{ab}$ gives
\begin{equation}
\dot{K} = 0
\quad\Rightarrow\quad
\dot{K}_{ab} = 2E_{ab}.
\end{equation}
In the strong–coupling limit the Gauss relation gives
\begin{equation}
E_{ab} = K K_{ab} - K_{a}{}^{c}K_{bc}.
\end{equation}
Taking the $n^a$-derivative and substituting $K^{\dot{}}=0$ and
$K^{\dot{}}_{ab}=2E_{ab}$, one finds exact cancellation of all terms, yielding
\begin{equation}
\dot{E}_{ab} = 0.
\end{equation}
Thus the staticity constraint follows from the evolution equation and the
scalar constraint in the ultralocal sector.

\subsection*{5. Shear constraint}

In the $1+3$ formalism with a perfect fluid, the Einstein--Bianchi system gives the algebraic shear constraint (Eq.~(\ref{Carroll:ELimit-Grav-05})):
\begin{equation}
-\tfrac12(\mu+p)\sigma_{ab}=0.
\end{equation}
\begin{itemize}
\item \textbf{Vacuum or $\Lambda$}: $\mu+p=0$, so the constraint is automatic.
\item \textbf{Silent perfect fluid}: $\mu+p\neq0$ forces $\sigma_{ab}=0$, i.e.\
      $K_{ab}$ becomes purely isotropic.
\end{itemize}
This constraint belongs to the matter sector; once matter is included and the
same silent limit is taken, it supplements the purely gravitational equations
derived from $S_E$.

In summary, in the electric sector, the scalar and evolution equations derived from the
electric Carroll action $S_E$ are equivalent, in the ultralocal/silent regime,s
to the full set of Carrollian electric equations Eq.~\ref{Carroll:ELimit-Grav-05} once the Bianchi
constraints and matter shear constraint are taken into account. In this sense
our electric Carrollian field equations are precisely the equations of motion
associated with the electric action of Ref.~\cite{Henneaux:2021gwp}, augmented by the requirement
of Carroll invariance.

In the magnetic sector, Ref.~\cite{Henneaux:2021gwp} only provides a Hamiltonian theory whose
equation of motion is $K_{ab}=0$, which freezes tensor degrees of freedom in
our FLRW/tensor setup. Our Carrollian magnetic equations \eqref{Carroll:HLimit-Grav-04}, derived by
imposing Carroll invariance on the $1+3$ field equations, instead describe a
nontrivial magneto-quasistatic gravitational regime with $E_{ab}=0$ but
$H_{ab}\neq 0$. They are therefore not the equations of motion of the magnetic
theory in Ref.~\cite{Henneaux:2021gwp}, but a complementary Carrollian limit that preserves
gravito-magnetic dynamics.

\input{Ref-2.bbl}
\end{document}

%% file: Ref-2.bbl
\providecommand{\href}[2]{#2}\begingroup\raggedright\endgroup